\documentclass[lettersize,journal]{IEEEtran}
\usepackage{amsmath,amsfonts}
\usepackage{algorithmic}
\usepackage{algorithm}
\usepackage{array}
\usepackage[caption=false,font=normalsize,labelfont=sf,textfont=sf]{subfig}
\usepackage{textcomp}
\usepackage{stfloats}
\usepackage{url}
\usepackage{verbatim}
\usepackage{graphicx}
\usepackage{cite}
\usepackage{xspace}
\usepackage{xcolor}
\usepackage{makecell}
\usepackage{array}
\usepackage{pgfplots}
\usepackage{pgfplotstable}
\usepackage{tikz}
\usepackage{caption}
\usepackage{subcaption}
\usepackage{booktabs}
\usepackage{multirow}
\usepackage{adjustbox}
\usepackage{hyperref}
\usepackage{amssymb} 
\begin{document}

\title{A Flow is a Stream of Packets: A Stream-Structured Data Approach for DDoS Detection}

\author{
Raja Giryes\thanks{R. Giryes, L. Shafir, and A. Wool belong to the School of Electrical Engineering, Tel Aviv University, ISRAEL. Emails: \emph{raja@tauex.tau.ac.il, lior.shafir@gmail.com, yash@eng.tau.ac.il}},
\and
Lior Shafir,
\and
Avishai Wool
}


\maketitle

\begin{abstract}
Distributed Denial of Service (DDoS) attacks are getting increasingly harmful to the Internet, showing no signs of slowing down. Developing an accurate detection mechanism to thwart DDoS attacks is still a big challenge due to the rich variety of these attacks and the emergence of new attack vectors. 
In this paper, we propose a new tree-based DDoS detection approach that operates on a flow as a \textit{stream} structure, rather than the traditional fixed-size record structure containing aggregated flow statistics. 
Although aggregated flow records have gained popularity over the past decade, providing an effective means for flow-based intrusion detection by inspecting only a fraction of the total traffic volume, they are inherently constrained. Their detection precision is limited not only by the lack of packet payloads, but also by their structure, which is unable to model fine-grained inter-packet relations, such as packet order and temporal relations. Additionally, inferring aggregated flow statistics must wait for the complete flow to end. Here we show that considering flow inputs as variable-length streams composed of their associated packet headers, allows for very accurate and fast detection of malicious flows. We evaluate our proposed strategy on the CICDDoS2019 and CICIDS2017 datasets, which contain a comprehensive variety of DDoS attacks. Our approach matches or exceeds existing machine learning techniques' accuracy, including state-of-the-art deep learning methods. Furthermore, our method achieves significantly earlier detection, e.g., with CICDDoS2019 detection based on the first 2 packets, which corresponds to an average time-saving of 99.79\% and uses only 4--6\% of the traffic volume.
\end{abstract}


\newcolumntype{L}[1]{>{\raggedright\let\newline\\\arraybackslash\hspace{0pt}}m{#1}}
\newcolumntype{C}[1]{>{\centering\let\newline\\\arraybackslash\hspace{0pt}}m{#1}}
\newcolumntype{R}[1]{>{\raggedleft\let\newline\\\arraybackslash\hspace{0pt}}m{#1}}

\definecolor{color1}{RGB}{54,197,239}
\definecolor{color2}{RGB}{223,30,90}
\definecolor{color3}{RGB}{45,182,124}
\definecolor{color4}{RGB}{235,177,45}
\definecolor{color5}{RGB}{74,21,75}
\definecolor{color10}{RGB}{44,160,44}

\section{Introduction}
Distributed Denial-of-Service (DDoS) attacks remain a major threat on the Internet. 
Since their first occurrence in 1996 \cite{07366989709452305}, the DDoS attacks keep evolving and growing in size, frequency and complexity \cite{Kaspersky2022,  Radware2022}. Recent advances in cloud services, processing technologies and the massive proliferation of IoT devices have created new attack opportunities, allowing attackers to spread malware on more infected devices than ever before \cite{roman2018mobile, vignau201910, xiao2018security}. These large infected botnets are used to launch destructive large-scale attacks, resulting in extreme outages of online services \cite{Radware2022}. For example, in 2016, a 1.5 Tbps DDoS attack was carried out over the infamous Mirai botnet (a vast army of hijacked internet-connected devices) \cite{antonakakis2017understanding}, knocking down several popular websites such as Twitter, Reddit, Netflix and many others.

Detection is a key component of thwarting DDoS attacks. Although considered a simple attack, DDoS refers to a rich family of attacks, all of which are designed to saturate the victim with a high amount of messages to exhaust its resources, disrupting its normal operation. These attacks are implemented on a variety of protocols such as UDP, TCP, DNS and HTTP, exploiting different resource limitations in the network, transport and application layers. 
During the last two decades, researchers have studied defense mechanisms against DDoS, and proposed several detection techniques, with a special attention in recent years given to a variety of Machine Learning (ML) tools that have been proposed to detect DDoS attacks \cite{arshi2020survey, al2020survey, mittal2022deep}.

These detection methods are often implemented in \textit{Network-based Intrusion Detection Systems} (NIDSs). NIDSs are primarily categorized into two types: \textit{Packet-based} and \textit{Flow-based}. Traditional NIDSs are packet-based, monitoring the entire packets' headers and payloads \cite{roesch1999snort, ficke2018characterizing}. This requires all network traffic to pass through the NIDS detection engine, forming a huge volume of data which may not be feasible to process in the ever-growing high-speed networks.
To address the high frequency of attacks, and the rise in volume of traffic, a new type of flow-based NIDS has emerged during the last decade \cite{umer2017flow, sperotto2010overview}. Flow-based NIDS inspects only collected flow information, which amounts to only a fraction of the whole traffic \cite{sperotto2010overview}.

A flow is defined by a stream of packets that share the same characteristics, i.e., for IPv4, a flow key is a five-tuple that includes the source IP address, destination IP address, source port, destination port and protocol values. Till now, existing methods have relied on aggregated flow information, i.e., a fixed size collection of features that contain the flow statistics.
Such flow data may include the start and end timestamps, as well as global flow statistics such as the total number of packets and bytes, minimum, maximum and mean packet lengths, or other data items required by the model that uses the flow (in our case, an attack detector).

While the use of flow records has been shown to detect attacks \cite{umer2017flow}, their characteristics impose inherent constraints. Flow records represent summarized information about network traffic, which negatively affects the discrimination power of detection methods \cite{umer2017flow, sperotto2011flow, elejla2018flow}. Flow-based detection precision is limited not only by the lack of packets' payloads, which limits the ability to perform deep packet inspection, but also lacks the fine-grained packet-level information such as temporal relations between packets (or subsets of packets) that constitute the flow. This means that the timing and sequence of events within a flow may not be fully captured. If a flow is short-lived or sporadic, the resolution of temporal details may be limited. 


\begin{figure}[t]	
\centering
\includegraphics[clip, trim=5.6cm 0cm 4.4cm 1.0cm,width=0.5\textwidth]{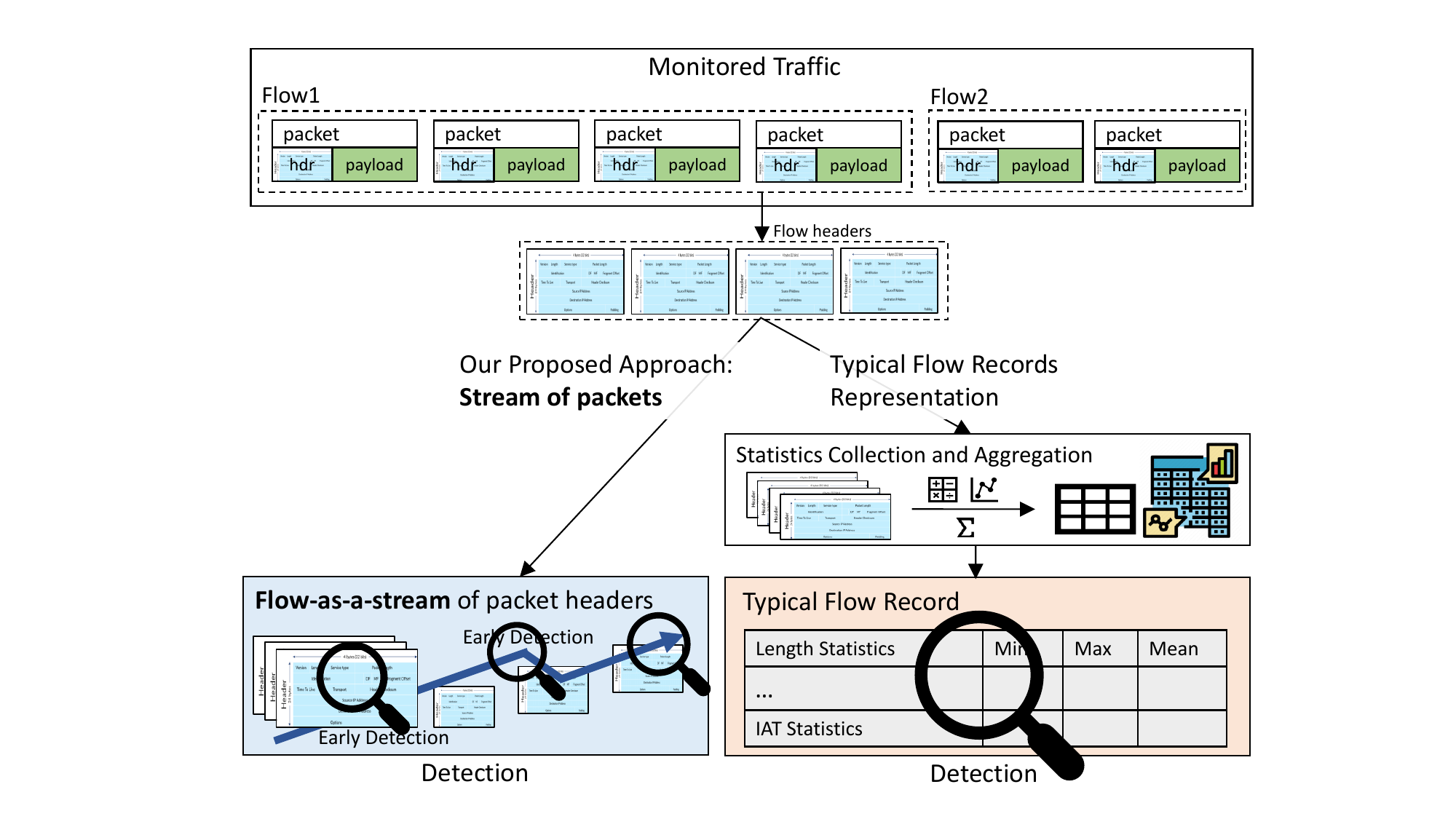}
\caption{\label{fig:our_method} An overview of our approach.} 
\end{figure}

Furthermore, aggregated flow records represent a snapshot of a communication session and are typically generated when a flow ends, forcing the detection system to \textit{wait for the completion of the flows}. Some DDoS attacks are designed to be low-and-slow \cite{pascoal2020slow}, meaning they generate traffic at a slow rate to evade rate-based detection mechanisms. These attacks can be challenging to identify as they do not exhibit the typical high-volume characteristics associated with traditional attacks.
Thus, to achieve a timely response, we believe that a real-time attack detection system in a modern live network must be able to process fractions of traffic flows partitioned across multiple time windows.

We propose a novel tree-based intrusion detection approach that considers a flow input as a \textit{stream} of packet headers, rather than a fixed-size record structure that contains aggregated flow statistics.
Most machine learning algorithms are not designed to operate on such variable-length stream structures and are limited to processing records of a fixed size. To handle such data, we employ a \textit{Set-Tree} model that is designed for variable length set data \cite{pmlr-v139-hirsch21a}. This technique enables the use of set data inputs in tree-based models, such as Random-Forest and Gradient-Boosting, by introducing an \textit{attention mechanism} and \textit{set-compatible split criteria}. 
The attention mechanism allows the decision tree to apply set-compatible operations to subsets of the input set, which were generated by a function in one of the nodes on the path leading to the current tree node.

Additionally, we show that the proposed stream input allows for early detection of malicious traffic. 
We train our detector on complete streams---however, at detection time the model is presented with just the first packets of a flow: we shall see that usually 2 or 4 packets suffice for very accurate detection. We shall see that learning the packet-level characteristics of the flow captures fine-grained but discernible information about the flow, information that is otherwise undetectable by traditional flow aggregation methods.

By analyzing the packet headers directly, we enjoy the fundamental benefit of capturing timing and sequence of events within the flow. Moreover, our detection method has additional advantages: (i) In traditional flow-based detection, which relies on aggregated flow records, there is an inherent overhead in collecting and generating the flow statistics as aggregated features. Our method does not require this step, as it operates on the raw data of packet headers .  
(ii) We use the same classifier for complete flows and early detection, based on a configurable number of first packets. 
(iii) Our stream-of-packets input is shown to enable learning of sub-flow (temporal packet-level) patterns. 
(iv) Similar to traditional flow-based detection, our model does not require packet payload inspection or processing. The flow representation includes only 15 features for each packet in the flow stream. 
(v) Our proposed method consumes a compact amount of data, using only 4--6\% of the total traffic volume. Note that in the majority of the DDoS attacks the flow size (in the number of packets) decreases by roughly 50\% \cite{ficke2018characterizing}.
(vi) A critical component of our tree-based model, in comparison to deep learning \cite{abeshu2018deep} and other complex methods, is its better interpretability, which may be achieved by measuring the frequency each packet (or subset of packets) occurs in the packets' attention-sets.

We evaluate our method on two data sets: the CICDDoS2019 dataset \cite{sharafaldin2019developing}, which contains a comprehensive variety of DDoS attack vectors; and the CICIDS2017 dataset, which includes various attack vectors, including some DDoS attacks.

We shall see that in terms of accuracy, our model outperforms existing statistical and ML (both classic and deep learning) methods that were evaluated over the CICDDoS2019 and CICIDS2017 datasets, including recent works described in \cite{sharafaldin2019developing, CIC1_hussain2020network, CIC2_ortet2021towards, CIC3_elsayed2020ddosnet, CIC4_li2020rtvd, CIC5_novaes2020long, CIC6_jia2020flowguard, CIC7_rajagopal2021towards, CIC8_almiani2021ddos, CIC9_de2020near, roopak2019deep, doriguzzi2020lucid, belarbi2022intrusion}. In addition to its high accuracy, our approach achieves a very rapid detection with an average time-saving of 99.79\% in terms of flows duration, while maintaining an accuracy over 0.999.

\section{Related Work} \label{sec:related_work}
\textbf{DDoS detection} has been studied by the research community during the last two decades.
Statistical and entropy-based approaches to anomaly detection in attacks have been proposed since the early 2000s. Feinstein et al. \cite{feinstein2003statistical} considered the use of entropy and frequency-sorted distributions of source IP addresses to identify DDoS attacks. The authors observed that DDoS traffic generated at that time, using non-sophisticated toolkits often has packet-crafting characteristics that make it possible to distinguish it from normal traffic. A similar approach is proposed by \cite{bojovic2019practical}, and is applied to both entropy and packet number time series. These entropy-based methods often rely on detection thresholds which require manual adjustments to adapt to different attack scenarios.

There has been extensive research on the use of machine learning algorithms in Network Intrusion
Detection Systems (NIDSs) for anomaly detection. Several classical machine learning algorithms were suggested in the NIDS literature including support vector machines (SVM) \cite{subbulakshmi2011detection}, Nearest Neighbor Classification \cite{ertoz2004minds}, Decision Trees \cite{qin2004frequent} and Random Forest \cite{chen2018detection}. However, 
the adoption of such solutions in NIDS is limited due to the high cost of errors, and the evolution of new network and application-layer attack vectors.


In \cite{chen2018xgboost}, an XGBoost classifier is used to detect DDoS attacks, while the authors of \cite{CIC9_de2020near} propose the use of Convolutional Neural Networks (CNNs) in IoT and SDN networks. The proliferation of insecure IoT devices has also created new opportunities for attackers to launch large-scale attacks using IoT botnets. Doshi et al. \cite{doshi2018machine} suggest that the unique characteristics of IoT devices (e.g., limited number
of endpoints and regular time intervals between packets) may result in high accuracy of intrusion detection in IoT network traffic using a variety of ML methods, including decision trees and deep learning.

\textbf{Datasets.} While some of the old classic DDoS attack vectors (e.g., TCP and SYN flood) remain an effective and major threat \cite{Kaspersky2022}, they have evolved in recent years to become stealthier, exposing less unique characteristics than the ones observed in the past. This, together with the emergence of new attack vectors (such as application-layer and complexity attacks) underline the need for rich, updated, and well-designed datasets, in order to evaluate new intrusion detection techniques, but also to evaluate the effectiveness of existing detection approaches. 

Most of the proposed intrusion detection techniques in the literature evaluate their work on a limited set of attack data. In some cases, simple tools such as Hping3 \cite{hping3}, Nmap and Scapy are used to simulate a SYN flood or simple UDP DDoS attacks, while in other cases, the evaluation is done on public datasets that are outdated or incomplete. In \cite{sharafaldin2018toward,sharafaldin2019developing}, the authors explored publicly available IDS and DDoS datasets spanning from 1999 to 2018, including DARPA98, CAIDA UCSD \cite{caida}, DARPA 2000, \cite{darpa}, Brown et al. \cite{brown2009analysis}, Singh and De \cite{singh2015approach}, Yu et al. \cite{yu2011discriminating}, and Subbulkashmi et al. \cite{subbulakshmi2011detection}, and found them outdated, non-heterogeneous, and having many issues and shortcomings, such as anonymized data, incomplete traffic, and lack of modern attack scenarios. To address the issues above, they provide new datasets named CICDDoS2019  \cite{sharafaldin2019developing}, and CICIDS2017 \cite{sharafaldin2018toward}, which we use in our work. They contain an updated and comprehensive set of intrusion and DDoS attacks.

\textbf{Detection Methods.} Since its publication, a variety of recent DDoS detection approaches have evaluated their work using the CICDDoS2019 dataset: Li et al. \cite{CIC4_li2020rtvd} propose a real-time entropy-based detection scheme using a sliding time window optimization. In another work, FlowGuard \cite{CIC6_jia2020flowguard} is proposed as an IoT defense scheme, employing two deep learning models: a long-short-term memory network (LSTM) and a CNN. Elsayed et al. \cite{CIC3_elsayed2020ddosnet} develop an SDN intrusion detection system (NIDS). The proposed system has two stages: (i) a pretraining stage uses an unsupervised auto-encoder (AE) model to produce a compressed representation of the input data; and (ii) supervised learning using a Recurrent Neural Network (RNN). The system reaches a high accuracy in comparison to existing detection methods. In \cite{CIC8_almiani2021ddos}, the authors use a deep Kalman back-propagation neural network to detect DDoS attacks in 5G-enabled IoT networks.
Note that all these approaches rely on the use of typical aggregated flow records. Conversely, in our work, the flow input representation is a variable-length stream, where each item in the stream contains packet header features.

Many works have also evaluated their techniques using the CICIDS2017 \cite{sharafaldin2018toward} dataset. Since CICIDS2017 includes various attack categories beyond DDoS, several works \cite{belarbi2022intrusion}, \cite{verkerken2023novel} propose multi-class detection models.
Here, we pay particular attention to binary-detection models which are similar to our work. 
In \cite{roopak2019deep}, the authors proposed four classifiers for detecting Distributed Denial of Service (DDoS) attacks using deep learning models, including MLP, 1D-CNN, LSTM, and 1D-CNN+LSTM. The experimental findings revealed that the CNN+LSTM model outperformed the other deep learning models, achieving an impressive accuracy of 97.16\%. In \cite{doriguzzi2020lucid}, a deep learning detection system for DDoS attacks is presented, called LUCID. The proposed model is a CNN-based IDS for binary classification of attacks, designed to be used in online resource-constrained environments. The evaluation results showed that the CNN-based IDS identifies DDoS attacks on the CICIDS2017 dataset with an accuracy of 99.7\%.

\textbf{Early-stage monitoring.} Several works have addressed the effectiveness of network classification at an early stage by monitoring the first few packets or bytes of a flow and employing supervised machine learning algorithms. 
Bernaille et al.~\cite{bernaille2006traffic} use the first $P$ packets' size and direction (during TCP connection setup) for early traffic classification. They represent each flow as a $P$-dimensional space, where the $p$'th coordinate is the size of the $p$'th packet, and employ k-Means clustering to determine the application class.
This approach reaches a medium-quality accuracy of about 80\% with $P=5$ and $k=50$ (number of clusters) for most application classes. However, a different application class that exhibits similar behavior during the application’s negotiation phase, can easily lead to misclassification of the flows. 
Also, TCP out-of-order packets that changes the spatial representation of the flow will impact the quality of the classification. 
Li et al.~\cite{li2008real} use C4.5 decision trees and REPTrees to perform early classification of Peer-to-Peer (P2P) applications based on the first few packets, and evaluated their work in comparison to the k-Means clustering method in \cite{bernaille2006traffic}.

Early classification of encrypted traffic is proposed by \cite{gu2011realtime,liu2017novel}. In \cite{gu2011realtime}, the authors use the C4.5, SVM, NB, and RF learning algorithms to classify the traffic based on packet length and inter-arrival time statistics of the first packets. A Window First N Packets (WFNP) is proposed by \cite{liu2017novel}. WFNP monitors the first N packets, extracts 11 statistical features, and deploys a C4.5 decision tree to classify the traffic. Both works \cite{gu2011realtime,liu2017novel}  evaluated their algorithms on private datasets.

While a variety of sub-flow sizes, selected features, and machine learning algorithms are evaluated in these early-stage classification proposals, they mostly focus on application traffic classification rather than on intrusion attacks and DDoS in particular, which by nature try to exhibit benign application traffic behavior. Moreover, our work differs from these works by considering a variable-length number of packets as an input. Our proposed classifier is not trained on a fixed number of packets, and may be presented with any sub-stream of first packets as an input during detection. Additionally, our proposed model is permutation-invariant (i.e., does not rely on items' order), thus it is robust against out-of-order packets that may occur in the network.

\section{Methodology} \label{sec:method}
We propose a new flow input stream structure capable of capturing inter-packet patterns, which in particular allows early detection of malicious flows. 
To understand the potential benefits of early detection, we analyze the datasets on which we evaluate our method, with a special focus on flows' duration and packet count across various attack vectors.  Since static datasets comprise complete flow records as labeled data, we reprocessed the associated .pcap raw data to reconstruct a packet-based labeled dataset, which we use for analysis, training and evaluation of our method.
Note that most machine learning algorithms are not designed to operate on stream structures and are limited to processing records of a fixed size. Our approach uses a \textit{Set-Tree} model that is designed for set data \cite{pmlr-v139-hirsch21a}.

In this section, we start by introducing the datasets that we use and their characteristics. Then, we describe the Set-Tree detection model, the inputs it uses, and how we reconstruct flow inputs as variable-sized streams of packets. We continue by presenting the packet header features that we use, and then discuss training and testing methodologies. 

\subsection{Datasets and Attack Vectors Characteristics} \label{subsec:dataset}
Our method is evaluated with recent datasets, CICDDoS2019 \cite{sharafaldin2019developing} and CICIDS2017 \cite{sharafaldin2018toward}, designed and provided by the Canadian Institute for Cybersecurity in collaboration with the University of New Brunswick.

\begin{table}
\caption{\label{table:flows-duration-2019} CICDDoS2019: Flows Duration Characteristics.}
\center
\footnotesize
\begin{tabular}{ m{1.4cm} | R{0.7cm} R{0.9cm} | R{1.8cm} R{1.8cm} }
\hline
  & \multicolumn{2}{c|}{Packets Count} & \multicolumn{2}{c}{Mean Duration (ms)} \\
\hline
Vector & Mean & Median & 2-packets & Complete flow \\
\hline
BENIGN & 12.96 & 4 & 20.777 & 8717.000 \\
NTP & 63.76 & 40 & 0.494 & 15.407 \\
TFTP & 3.42 & 4 & 0.062 & 2010.000 \\
SNMP & 3.48 & 2 & 0.006 & 1.280 \\
LDAP & 4.45 & 2 & 0.007 & 0.467 \\
NetBIOS & 2.01 & 2 & 0.010 & 28.692 \\
MSSQL & 2.09 & 2 & 0.006 & 6.374 \\
Syn & 3.26 & 2 & 0.131 & 8088.000 \\
UDP & 3.75 & 3 & 0.011 & 89.865 \\
SSDP & 3.71 & 2 & 0.008 & 82.636 \\
UDP-LAG & 3.16 & 2 & 0.049 & 5222.000 \\
DNS & 5.92 & 2 & 0.068 & 1.203 \\
\hline 
Overall & 8.66 & 4 & 10.391 & 4918.738 \\
\end{tabular}
\label{tab:ddos_duration_table}
\end{table}

\begin{table}
\caption{\label{table:flows-duration-2017} CICIDS2017: Flows Duration Characteristics.}
\center
\footnotesize
\begin{tabular}{ m{1.5cm} | R{0.7cm} R{0.7cm} | R{0.75cm} R{0.75cm} R{0.75cm} R{0.75cm} }
\hline
  & \multicolumn{2}{c|}{Packets Count} & \multicolumn{4}{c}{Mean Duration (ms)} \\
\hline
Vector & Mean &  Median & 2-packets & 4-packets & 14-packets& Complete flow\\
\hline
BENIGN & 22.74 & 4 & 141 & 1565 & 3455 & 11092\\
GoldenEye & 9.17 & 10 & 11900 & 11968 & 18786 & 19408\\
Slowloris & 7.99 & 4 & 470 & 22320 & 34949& 56554 \\
DoS Hulk & 9.47 & 12 & 1521 & 16895& 55599 & 56044 \\
Slowhttptest & 6.74 & 7 & 1131 & 5602 & 55492 & 58615 \\
Brute Force & 18.42 & 4 & 0.116 & 4724 & 5033 & 6502 \\
Web XSS & 11.01 & 4 & 0.128 & 5101 & 5184 & 6612 \\
PortScan & 2.02 & 2 & 0.060 & 80 & 81 & 82 \\
Bot & 6.54 & 6 & 0.503 & 126 & 328 & 328 \\
DDoS & 7.90 & 8 & 4.8 & 1099 & 15352 & 15837 \\
FTP-Patator & 12.86 & 12 & 14.7 & 19 & 1540 & 4513 \\
SSH-Patator & 26.91 & 26 & 0.752 & 2.5 & 2200 & 6168 \\
\hline 
Overall & 14.57 & 4 & 547 & 4758 & 15465 & 19277 \\
\end{tabular}
\label{tab:ids_duration_table}
\end{table}

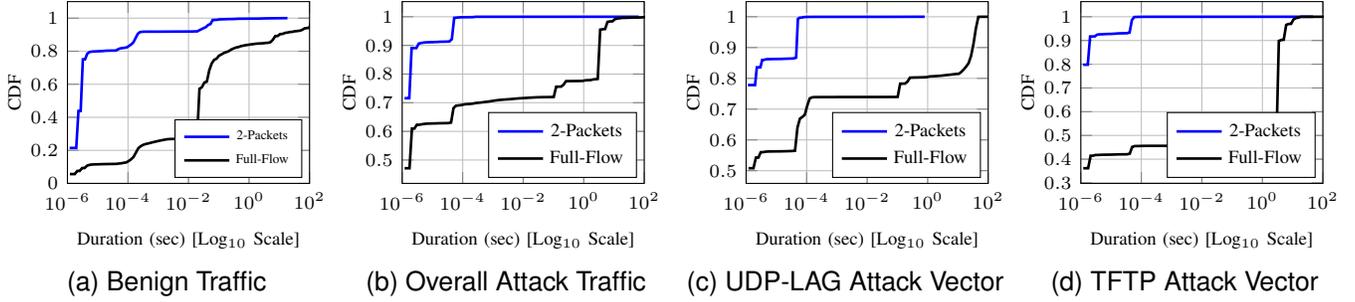
\begin{figure*}[!t]
  \centering
  \subfloat[Benign Traffic]{%
    \begin{tikzpicture}
        \begin{axis}[xlabel={Duration (sec) [Log$_{10}$ Scale]},
            height=4cm,
            width=4.8cm,
            ylabel={CDF},
            ylabel style={yshift=-15pt},
            ylabel style={font=\scriptsize},
            yticklabel style={font=\scriptsize}, 
            xlabel style={font=\scriptsize}, 
            xticklabel style={font=\scriptsize}, 
            xtick={-6,-4,-2, 0, 2},
            xticklabels={$10^{-6}$, $10^{-4}$, $10^{-2}$, $10^{0}$, $10^{2}$},
            xmin=-6,
            xmax=2,
            ymin=0,
            grid=both,
            ytick distance=0.2,
            legend style={font=\tiny},
            legend pos=south east,
            ]
        \addplot[line width=1pt, blue] table[x index=0,y index=1,col sep=comma] {csv/ddos/cdf/BENIGN-2packet-flow-log.csv};  
        \addplot[line width=1pt] table[x index=0,y index=1,col sep=comma] {csv/ddos/cdf/BENIGN-full-flow-log.csv}; 
        \legend{2-Packets, Full-Flow}
        \end{axis}
    \end{tikzpicture}%
  }
  \subfloat[Overall Attack Traffic]{%
    \begin{tikzpicture}
              \begin{axis}[xlabel={Duration (sec) [Log$_{10}$ Scale]},
            height=4cm,
            width=4.8cm,
            ylabel={CDF},
            ylabel style={yshift=-15pt},
            ylabel style={font=\scriptsize},
            yticklabel style={font=\scriptsize}, 
            xlabel style={font=\scriptsize}, 
            xticklabel style={font=\scriptsize},
            xtick={-6,-4,-2, 0, 2},
            xticklabels={$10^{-6}$, $10^{-4}$, $10^{-2}$, $10^{0}$, $10^{2}$},
            xmin=-6,
            xmax=2,
            ytick distance=0.1,
            grid=both,
            legend style={font=\scriptsize},
            legend pos=south east,
            ]
        \addplot[line width=1pt, blue] table[x index=0,y index=1,col sep=comma] {csv/ddos/cdf/overallattack-2packet-flow-log.csv};  
        \addplot[line width=1pt] table[x index=0,y index=1,col sep=comma] {csv/ddos/cdf/overallattack-full-flow-log.csv}; 
        \legend{2-Packets, Full-Flow}
        \end{axis}
    \end{tikzpicture}
  }
  \subfloat[UDP-LAG Attack Vector]{%
    \begin{tikzpicture}
        \begin{axis}[    xlabel={Duration (sec) [Log$_{10}$ Scale]},
            height=4cm,
            width=4.8cm,
            ylabel={CDF},
            ylabel style={yshift=-15pt},
            ylabel style={font=\scriptsize},
            yticklabel style={font=\scriptsize}, 
            xlabel style={font=\scriptsize}, 
            xticklabel style={font=\scriptsize}, 
            xtick={-6,-4,-2, 0, 2},
            xticklabels={$10^{-6}$, $10^{-4}$, $10^{-2}$, $10^{0}$, $10^{2}$},
            xmin=-6,
            ytick distance=0.1,
            xmax=2,
            grid=both,
            legend style={font=\scriptsize},
            legend pos=south east,
            ]
        \addplot[line width=1pt, blue] table[x index=0,y index=1,col sep=comma] {csv/ddos/cdf/UDP-lag_2packet_flow_log.csv};  
        \addplot[line width=1pt] table[x index=0,y index=1,col sep=comma] {csv/ddos/cdf/UDP-lag_full_flow_log.csv}; 
        \legend{2-Packets, Full-Flow}
        \end{axis}
        \end{tikzpicture}}
  \subfloat[TFTP Attack Vector]{%
    \begin{tikzpicture}
        \begin{axis}[    xlabel={Duration (sec) [Log$_{10}$ Scale]},
            height=4cm,
            width=4.8cm,
            ylabel={CDF},
            ylabel style={yshift=-15pt},
            ylabel style={font=\scriptsize},
            yticklabel style={font=\scriptsize}, 
            xlabel style={font=\scriptsize}, 
            xticklabel style={font=\scriptsize},
            xtick={-6,-4,-2, 0, 2},
            ytick distance=0.1,
            xticklabels={$10^{-6}$, $10^{-4}$, $10^{-2}$, $10^{0}$, $10^{2}$},
            xmin=-6,
            xmax=2,
            grid=both,
            legend style={font=\scriptsize},
            legend pos=south east,
            ]
        \addplot[blue, line width=1pt] table[x index=0,y index=1,col sep=comma] {csv/ddos/cdf/TFTP_2packet_flow_log.csv};  
        \addplot[line width=1pt] table[x index=0,y index=1,col sep=comma] {csv/ddos/cdf/TFTP_full_flow_log.csv}; 
        \legend{2-Packets, Full-Flow}
        \end{axis}
    \end{tikzpicture}
  }
  \caption{CICDDoS2019 flow duration distribution - a comparison between first 2-packets and complete flows.}
  \label{fig:twosidebyside}
\end{figure*}

\begin{figure*}[t]
  \centering
  \subfloat[Benign Traffic]%
    {\begin{tikzpicture}
        \begin{axis}[xlabel={Duration (sec) [Log$_{10}$ Scale]},
            height=4cm,
            width=4.8cm,
            ylabel={CDF},
            ylabel style={yshift=-15pt},
            ylabel style={font=\scriptsize},
            yticklabel style={font=\scriptsize}, 
            ytick distance=0.2,
            xlabel style={font=\scriptsize}, 
            xticklabel style={font=\scriptsize}, 
            xtick={-6,-4,-2, 0, 2},
            xticklabels={$10^{-6}$, $10^{-4}$, $10^{-2}$, $10^{0}$, $10^{2}$},
            xmin=-6,
            xmax=2,
            grid=both, 
            ]
        \addplot[color10, line width=1pt, name=BENIGN_2] table[x index=0,y index=1,col sep=comma] {csv/ids/cdf/BENIGN_2packet_flow_log.csv}; 
        \addplot[blue, line width=1pt, name=BENIGN_4] table[x index=0,y index=1,col sep=comma] {csv/ids/cdf/BENIGN_4packet_flow_log.csv}; 
        \addplot[color2, line width=1pt, name=BENIGN_f] table[x index=0,y index=1,col sep=comma] {csv/ids/cdf/BENIGN_full_flow_log.csv}; 
        \end{axis}
    \end{tikzpicture}
    }
   \subfloat[Overall Attack Traffic]{%
      \begin{tikzpicture}
        \begin{axis}[xlabel={Duration (sec) [Log$_{10}$ Scale]},
            height=4cm,
            width=4.8cm,
            ylabel={CDF},
            ylabel style={yshift=-15pt},
            ylabel style={font=\scriptsize},
            yticklabel style={font=\scriptsize}, 
            xlabel style={font=\scriptsize}, 
            xticklabel style={font=\scriptsize},
            xtick={-6,-4,-2, 0, 2},
            xticklabels={$10^{-6}$, $10^{-4}$, $10^{-2}$, $10^{0}$, $10^{2}$},
            xmin=-6,
            xmax=2,
            ytick distance=0.2,
            grid=both,
            ]
        \addplot[color10, line width=1pt] table[x index=0,y index=1,col sep=comma] {csv/ids/cdf/overallattack_2packet_flow_log.csv};  
        \addplot[blue, line width=1pt] table[x index=0,y index=1,col sep=comma] {csv/ids/cdf/overallattack_4packet_flow_log.csv}; 
        \addplot[color2, line width=1pt] table[x index=0,y index=1,col sep=comma] {csv/ids/cdf/overallattack_full_flow_log.csv}; 
        \end{axis}
        \end{tikzpicture} 
    }
    \subfloat[DoS Hulk Vector]{%
      \begin{tikzpicture}
        \begin{axis}[    xlabel={Duration (sec) [Log$_{10}$ Scale]},
            height=4cm,
            width=4.8cm,
            ylabel={CDF},
            ylabel style={yshift=-15pt},
            ylabel style={font=\scriptsize},
            yticklabel style={font=\scriptsize}, 
            xlabel style={font=\scriptsize}, 
            xticklabel style={font=\scriptsize}, 
            xtick={-6,-4,-2, 0, 2},
            xticklabels={$10^{-6}$, $10^{-4}$, $10^{-2}$, $10^{0}$, $10^{2}$},
            xmin=-6,
            xmax=2,
            ytick distance=0.2,
            grid=both,
            ]
        \addplot[color10, line width=1pt] table[x index=0,y index=1,col sep=comma] {csv/ids/cdf/DoS_Hulk_2packet_flow_log.csv};  
        \addplot[blue, line width=1pt] table[x index=0,y index=1,col sep=comma] {csv/ids/cdf/DoS_Hulk_4packet_flow_log.csv}; 
        \addplot[color2, line width=1pt] table[x index=0,y index=1,col sep=comma] {csv/ids/cdf/DoS_Hulk_full_flow_log.csv}; 
        \end{axis}
        \end{tikzpicture} 
        }
    \subfloat[DoS GoldenEye Vector]{%
        \begin{tikzpicture}
        \begin{axis}[    xlabel={Duration (sec) [Log$_{10}$ Scale]},
            height=4cm,
            width=4.8cm,
            ylabel={CDF},
            ylabel style={yshift=-15pt},
            ylabel style={font=\scriptsize},
            yticklabel style={font=\scriptsize}, 
            xlabel style={font=\scriptsize}, 
            xticklabel style={font=\scriptsize},
            xtick={-6,-4,-2, 0, 2},
            xticklabels={$10^{-6}$, $10^{-4}$, $10^{-2}$, $10^{0}$, $10^{2}$},
            xmin=-6,
            xmax=2,
            ytick distance=0.2,
            grid=both, 
            ]
        \addplot[color10, line width=1pt] table[x index=0,y index=1,col sep=comma] {csv/ids/cdf/DoS_GoldenEye_2packet_flow_log.csv};  
        \addplot[blue, line width=1pt] table[x index=0,y index=1,col sep=comma] {csv/ids/cdf/DoS_GoldenEye_4packet_flow_log.csv}; 
        \addplot[color2, line width=1pt] table[x index=0,y index=1,col sep=comma] {csv/ids/cdf/DoS_GoldenEye_full_flow_log.csv}; 
        \end{axis}
        \end{tikzpicture} 
        }
        \hfil
\hspace{1cm}
\begin{tikzpicture}
  \matrix[name=mylegend,anchor=south, draw=black, thin, row sep=0.5em] at (0,0) {
    \node [label=right:{\footnotesize First 2-Packets}] {\protect\raisebox{-0.5ex}{\color{green}\rule{1em}{1pt}}}; & 
    \node [label=right:{\footnotesize First 4-Packets}] {\protect\raisebox{-0.5ex}{\color{blue}\rule{1em}{1pt}}}; & 
    \node [label=right:{\footnotesize Complete-Flow}] {\protect\raisebox{-0.5ex}{\color{red}\rule{1em}{1pt}}}; \\
  };
\end{tikzpicture}
\hspace{1cm}
    \caption{CICIDS2017 flow duration distribution - a comparison between initial-packet streams and complete flows.}
  \label{fig:cic_ids_cdf_figure}
\end{figure*}
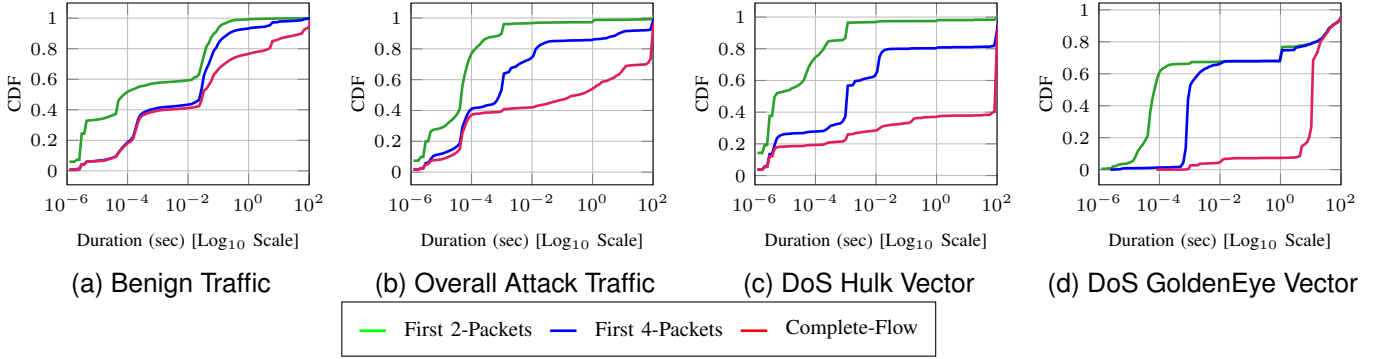

CICDDoS2019 contains more than 50 million DDoS attack flows including UDP, SNMP, NetBIOS, LDAP, TFTP, NTP, SYN, MSSQL, UDP-Lag, DNS and SSDP DDoS attack vectors, and more than 100 thousand flows of benign traffic, collected in two separate days.

CICIDS2017 dataset contains both benign and modern common attacks including Brute Force FTP, Brute Force SSH, DoS, Heartbleed, Web Attack, Infiltration, Botnet and DDoS. To generate realistic benign background traffic, the dataset authors profiled the abstract behaviour of human interactions of 25 users based on the HTTP, HTTPS, FTP, SSH and email protocols. 
The collected traffic in both datasets is available in raw captured files (PCAP) and in labeled flow-based CSV format. The flow-based format includes more than 80 features for each labeled flow.




\subsubsection{Flows Duration in CICDDoS2019} To quantify the benefits of early-detection, we analyzed the packet count and duration of flows in both datasets. 
The number of packets and mean duration of flows across CICDDoS2019 benign and attack traffic are presented in Table \ref{tab:ddos_duration_table}. We compare the mean duration of complete flows and the mean duration of the first two packets in each flow. The focus on the first two packets is based on the relatively low median and mean packet count values of DDoS attack flows in CICDDoS2019. Importantly, although the median packet counts are relatively low, early detection based on the first two packets offers a significant time saving of 99.79\%. This is underscored by the fact that while the mean duration of complete flows in CICDDoS2019 is 4.91 seconds, the mean duration of the first two packets is only 10.39 \emph{milliseconds} (ms). 
The cumulative distribution of benign and attack flow durations in CICDDoS2019 is depicted in Figures \ref{fig:twosidebyside}a and \ref{fig:twosidebyside}b, respectively. Note that more than 30.6\% of the attack flows complete durations are above 0.1 ms, while 99.6\% of the same flows exhibit durations below 0.0525 ms for their first two packets. In the case of benign traffic, 73\% of the flows have durations exceeding 10 ms, while 83\% of the corresponding first two packets' flow durations are below 0.1 ms. 
Additionally, as examples we present the duration distributions of two specific DDoS attack vectors, UDP-LAG and TFTP, in Figures \ref{fig:cic_ids_cdf_figure}c and \ref{fig:cic_ids_cdf_figure}d respectively. The duration distributions of all the attack vectors are provided in Appendix \ref{appendix:duration_attack_vectors}.

\textbf{Example: TFTP Attack flow durations}. In the case of the TFTP attack, the mean flow duration is 2.01 seconds, while the mean duration of the first two packets is only 0.062 ms (See Table \ref{tab:ddos_duration_table}). The TFTP attack, which stands for Trivial File Transfer Protocol Distributed Attack, exploits the asymmetry between request and response sizes, leading to amplification. In this attack, the attacker spoofs the victim's source IP address and targets TFTP servers for reflection. Analysis of the CICDDoS2019 dataset reveals that out of the 20,043,067 TFTP attack flows, 9,432,814 flows consist of two packets, representing consecutive UDP requests with an average elapsed time of 0.017 ms between them. However, 7,961,980 flows consist of four packets, representing two pairs of two consecutive UDP requests, where the average elapsed time between the pairs is 3 seconds. Thus being able to accurately detect the attack after the first two packets drastically reduces detection time.

\subsubsection{Flows Duration in CICIDS2017} Table \ref{tab:ids_duration_table} presents the packet count and duration characteristics of CICIDS2017 benign and attack traffic. Unlike CICDDoS2019, this dataset includes various attack vectors beyond DDoS, resulting in higher values of packet counts and flow durations. Therefore, in addition to the first two packets, we provide duration characteristics for the first four and fourteen packets as well. The potential time savings in this dataset are somewhat lower than in the CICDDoS2019 dataset. The mean duration of complete flows in CICIDS2017 is 19.27 seconds, while the mean durations of the first two and four packets are 546.76 ms (-97.16\%) and 4.75 seconds (-75.31\%), respectively. The cumulative distributions of benign, all attacks, DoS Hulk and GoldenEye durations in CICIDS2017 are depicted in Figure \ref{fig:cic_ids_cdf_figure}. Note that more than 45\% of the attack complete flow durations are above 1 second, while 90\% of the corresponding first two packets' flow durations are below 1 ms.

The characteristics of duration and packet count across the CICDDoS2019 and CICIDS2017 datasets highlight the potential time-saving benefits of early detection of different packet patterns and attack vectors.

\subsection{Set Tree Detection Model} \label{subsec:set-tree-model}
As our flow input data is a set, we employ the discriminative Set-Tree method \cite{pmlr-v139-hirsch21a}, which supports using sets in tree-based models (vanilla decision trees cannot handle sets).  
Note that we select a tree-based approach as these still exhibit superiority on tabular data compared to deep learning techniques \cite{NEURIPS2022_0378c769}.

As in many other applications where a single sample represents a set of items, a network flow is defined as a stream of packets.  A prevalent practice used to address this gap is feature augmentation, e.g., aggregation of minimum, maximum, sum, mean etc.\ for each packet feature in the flow. As shown by \cite{pmlr-v139-hirsch21a}, feature augmentation alone might degrade the descriptive power of the decision tree. Therefore, we treat the packet-level data directly.

As an illustrative example, consider the case where $F$ is a flow that is composed of a stream of packets and the function $f$ is True (e.g., an attack) if the number of forward packets with length higher than 70 bytes is at least 5, and False otherwise. To evaluate $f$ using a vanilla decision tree, there should be a dedicated augmented feature that counts the number of packets that conform to this particular criterion, which is infeasible for an ever-growing amount of such functions required to accurately match different patterns observed in real-world traffic. However, such a criterion $f$ is easily described by a Set-Tree model.

Therefore, we suggest using the set-tree model. We start by briefly describing this model and its attention mechanism that addresses the issues described above. 

\begin{figure}[t]	
\centering
\includegraphics[clip, trim=10.6cm 9.6cm -16.2cm 2.5cm,width=1\textwidth]{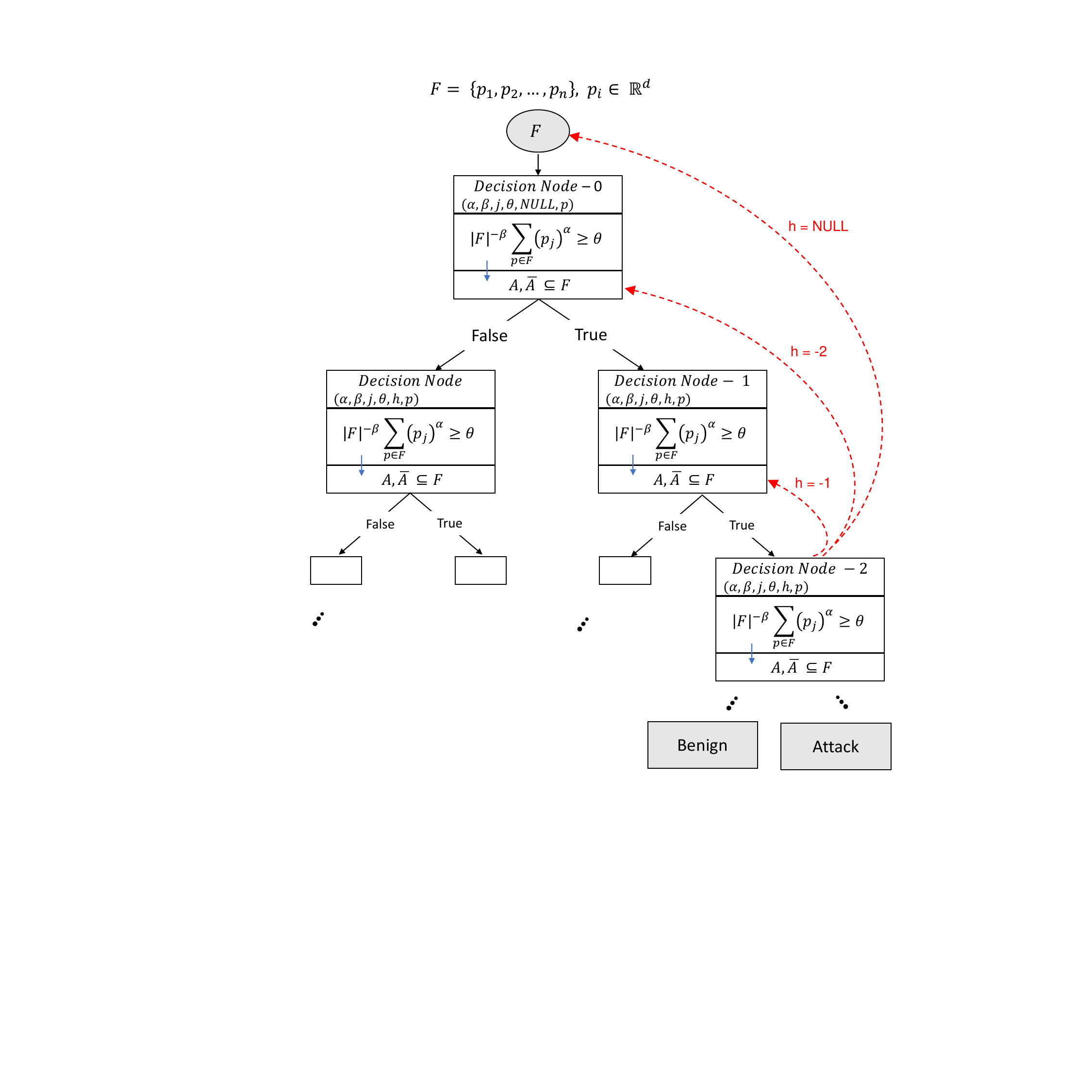}
\caption{\label{fig:set_tree}Using Set Tree \cite{pmlr-v139-hirsch21a} for network flow detection. Each decision node contains a set-compatible split criterion, and returns a subset of its items (packets attention set). The dashed arrows demonstrate the attention mechanism, where each decision node operates on the original input set $F$, or on one of the attention sets returned by previous nodes along the decision path.} 
\end{figure}

\textbf{Set-Tree set-compatible split criterion}. In most implementations of decision trees, the split criterion at each node of the tree is a binary function that operates on a single feature of the input record. In a Set-Tree, the split criterion is extended to a set function that operates on a certain feature in all items in the node's input set. More formally, consider a flow $F = {p_1, p_2, .., p_n}$, s.t.\ $p_i\in \mathbb{R}^d$ (each packet has $d$ features), the set decision function is defined as ${|F|}^{-\beta}\sum_{p\in F} (p_j)^\alpha \geq \theta$ where $j$ represents the index of the packet's feature to be used for the evaluation and $\theta$ is the threshold, $\alpha\in[-\infty,\infty]$, and $\beta\in\{0,1\}$ dictates whether to normalize by the size of the flow stream. Note that the evaluation function is defined for sets of any finite size, the number of items in a set may vary and predictions are expected to be independent of the items' ordering (i.e., permutation-invariant). The criterion function generalizes several statistical functions, e.g., $\alpha=\beta=0$ is the set sum function, $\alpha=\beta=1$ is the average over all packets' $j$'s feature, and with $\alpha=-1, \beta=1$ the inverse of the harmonic mean is used.

\textbf{Attention mechanism}.  
For each decision node, the set-compatible evaluation function induces an attention function $A$ that returns a subset of items which contains the most significant items in the input set that triggered the evaluation function to be True. More formally, given an input set of packets $F$, $A$ is defined as the following subset of packets:
\[A(F)=\{p \in F : (p_j)^\alpha \geq \frac{\theta} {|F|^{1-\beta}} \}\]
$\overline{A}$ is the complementary set to $A$. The attention mechanism allows each decision node to operate on the original input set, or on one of the attention sets $\overline{A}$, $A$ that were returned by the decision nodes in the path leading to the current node. Figure \ref{fig:set_tree} exhibits the Set-Tree where each node has a set-compatible evaluation function. The attention mechanism is depicted by the dashed arrows, and the corresponding parameter $h$, which may point to a preceding node (or $NULL$ in case that the original input set should be processed), $p$ is a flag that indicates whether the pointed $A$ attention set should be used, or the complementary set $\overline{A}$. The attention mechanism may express a dependency between the decision nodes. Consider the example described at the beginning of this section: if $A_1$ is the subset of forward packets within the input flow, $A_2$ is the subset of packets with length greater than 70, and $A_3$ is the attention set of a function that evaluates to True if the number of packets is greater than or equal to 5, then $A_3(A_2(A_1(F)))$ represents the required evaluation function.

\begin{table}[t]
\caption{\label{table:packet-features} Packet header features used in our model}
\small
\center
\begin{tabular}{ l | m{5.4cm}}
\hline
\textbf{Feature Name} & \textbf{Description}\\
\hline
Is forward & Packet's direction indication (ingoing or outgoing)\\
\hline
Source Port &  If less or equal to 1023, otherwise 0\\
\hline
Destination Port &  If less or equal to 1023, otherwise 0\\
\hline
Index & Packet's order in the flow stream\\
\hline
Timestamp &  Relative timestamp of the packet in the flow\\
\hline
Protocol & Packet's protocol header field \\
\hline
Length & Length of the packet in bytes \\
\hline
Init win bytes &  Number of bytes in initial window\\
\hline
IAT before & Time since the previous packet in the flow \\
\hline
IAT after & Time until the next packet in the flow \\
\hline
IAT before direction & Time since the previous packet in the same direction in the flow \\
\hline
IAT after direction & Time since the next packet in the same direction in the flow  \\
\hline
Is SYN & Is SYN flag set \\
\hline
Is ACK & Is ACK flag set  \\
\hline
Is RST & Is RST flag set  \\
\hline
\end{tabular}
\end{table}

\subsection{Data processing} \label{subsec:dataprocess}
To bridge the gap between the flow-level labeled information in the datasets and the packet-level header data required by our model, we performed the reprocessing of the raw captured PCAP files. This involved extracting packet header information and reconstructing the flows in a stream-of-packets structure.

To establish the correspondence between the raw packets and the labeled flows, we developed a packet-sniffing and processing tool specifically designed for the CICDDoS2019 and CICIDS2017 datasets. This tool allowed us to process the entire raw PCAP data and extract relevant flow information. The matching process relied on the flow key, which consists of the IPv4 five tuple: source IP address, destination IP address, source port, destination port, and protocol. Additionally, the start timestamp of each flow was used to disambiguate cases where flow keys were not unique. For every matched flow, we constructed an ordered-set structure (stream), where each item in the stream represents a packet in the flow. Each packet is represented as a fixed size collection of packet's header features. 
We will release an implementation of our data reprocessing method coupled with the packet-level labeled version of these datasets for the community and for reproducibility.

\textbf{Packet's Header Feature Selection.} The task of feature selection is critical in the construction and design of any discriminative model.  In our proposed model, the features are obtained from individual packets, with the promise of encoding flow-independent characteristics. On the one hand, our model's flow representation does not require to maintain flow information as features, i.e., aggregating statistics over multiple packets, over time, as long as the flow traffic evolves. On the other hand, as described in Section \ref{subsec:set-tree-model}, the tree nodes' split criterion is designed to compute statistical functions on a set of packet features. Thus, it is important to encode and select features that, once composed, have the potential of expressing flow stream or sub-flow stream characteristics. The features we use in our model are presented in Table \ref{table:packet-features}. 
Here we provide more information about the feature selection considerations:
\begin{table*}[t]
\caption{Comparison between the flow input data size of typical aggregated flow records and  streams of packet headers}
\center
\small
\begin{tabular}{ p{3cm}| p{2.4cm} | p{1.3cm} p{2.1cm} p{1.1cm} | p{1.3cm} p{2.1cm} p{1.1cm}}
\hline
  & Traffic & \multicolumn{3}{c|}{Flows} & \multicolumn{3}{c}{Packets} \\
\hline
Dataset & \footnotesize{Raw PCAP Size (GB)} & \footnotesize{\#Flows} \footnotesize{(Millions)} &\footnotesize{NetFlow Records Size(GB)} & \footnotesize{Traffic Portion} & \footnotesize{\#Packets} \footnotesize{(Millions)} & \footnotesize{Headers Size(GB)} & \footnotesize{Traffic Portion}\\
\hline
DDoS2019 & \multicolumn{1}{r|}{163.6} & \multicolumn{1}{r}{50.0}
 & \multicolumn{1}{r}{7.0} & 4.28\% & \multicolumn{1}{r}{248.8}
 & \multicolumn{1}{r}{9.3} & 5.68\% \\
IDS2017 & \multicolumn{1}{r|}{52.3} & \multicolumn{1}{r}{2.8}
 & \multicolumn{1}{r}{0.4} & 0.76\% & \multicolumn{1}{r}{55.6}
 & \multicolumn{1}{r}{2.0} & 4.02\%\\
\hline 
Overall & \multicolumn{1}{r|}{215.9}
 & \multicolumn{1}{r}{52.8}
 & \multicolumn{1}{r}{7.4} & 3.43\%  & \multicolumn{1}{r}{304.5}
 & \multicolumn{1}{r}{11.3} & 5.23\% \\
 \hline 
\end{tabular}
\label{table:data-size-comparison}
\end{table*}
 
\textbf{Overfitting.} We find that the usage of IP socket features, such as source IP address, destination IP address, and absolute timestamps do not add additional information to the detection problem, and cause the model to overfit. Models that rely on IP addresses, or other concrete socket information of the flow may perform well only in specific crafted attack scenarios, or become dependent on the setup on which the attack was performed. Thus, we do not include IP address information, and encode only source and destination port values that are lower or equal to 1023 (this range represents the well-known ports, that are reserved for common TCP/IP applications). Other port values (greater than 1023) are marked as zero. As for the timestamp, we normalize it to the flow start timestamp, i.e., we encode the packet timestamp as the time passed since the beginning of the flow.

\textbf{Index.} Although the Set-Tree model handles set structures, which are permutation-invariant (i.e., predictions are independent of the packets' ordering within the stream structure), for each packet we encode its index in the flow. This allows the model to learn patterns from subsets of packets before or beyond a certain index in the flow.

\textbf{Is Forward.} For each packet, we mark its direction as forward or backward. Naturally, some attack characteristics may be reflected in one direction of the traffic (e.g., from the attacker to the target). Therefore, we would like the model to operate on subsets of forward or backward packets. Note that the majority of the flow features (48 out of 81) provided in the datasets that we use, include separate statistics per direction, e.g., Fwd Packet Length Mean/Min/Max/Std, and Bwd Packet Length Mean/Min/Max/Std, etc.

\textbf{IAT - inter arrival time.} IAT is defined as the amount of time that elapses after the receipt of a packet until the next packet arrives. Here, for each packet we encode four IAT values, one for the time elapsed since the previous packet, and one for the time elapsed until the next packet. Two additional IAT values are encoded for the previous and next packets in the same direction (See Table \ref{table:packet-features}). Technically, this imposes the same information to be collected twice, i.e., the time elapsed between $p_i$ and $p_{i+1}$ is saved as a feature for both packets. However, since a packet may appear in an attention set without its previous or subsequent packet, we decide to keep this information individually per packet. According to \cite{doshi2018machine}, the vast majority of DDoS attack traffic has close to zero inter-packet arrivals and high first and second derivatives of inter-packet intervals.

\subsection{Inspected-Data Volume} \label{subsec:datasize}
The flow representation used in our method is a varying length stream of packet features, rather than a fixed size flow record. Here we quantify the proportion of the entire traffic is used by our detection method and compare it to the usage of typical aggregated flow records. 
To this end, we examine the number of flows and packets in both the CICDDoS2019 and CICIDS2017 datasets, as presented in Table \ref{table:data-size-comparison}.

The second column in the table represents the total size of the raw PCAP files, reflecting the overall traffic volume of the datasets. 
To estimate the size of typical flow records, we choose the NetFlow format \cite{NetFlow}, a widely deployed protocol for network traffic collection, commonly used by researchers and network administrators to monitor the network usage. For our analysis, we consider a flow record size of 150 bytes, although in general it may vary, depending on the configuration and number of fields included. Clearly, this is just a parameter of our method and it can support other sizes.
To determine the data size consumed by our method, we take into account the sizes of the IP header (20 bytes) and the TCP header (20 bytes), which together contribute to a total of 40 bytes per packet. 
Note that we assume a worst-case here, where all packets are carried over TCP (UDP header size is 8 bytes). Additionally, our method uses only part of the fields in these headers, as appeared earlier in Table \ref{table:packet-features}.

DDoS2019 contains much more attack flows than benign flows, while IDS2017 contains much more benign flows than attack flows. Thus, the ratio between the number of flows and number of packets varies across the two datasets. Table \ref{table:data-size-comparison} shows that Netflow-based flow records use 1-5\% of the traffic volume, while streams of packet headers use 4-6\% of the traffic volume.

\subsection{Training and Testing Methodology} \label{subsec:training}
We consider a binary classification model to classify the input flow as Benign or Attack. More formally,
\[
    f(y)= 
\begin{cases}
    \text{Benign},& \text{if } y = \text{Benign}\\
    \text{Attack}, & \text{otherwise},
\end{cases}
\]
where $y$ denotes the label of the flow input.

Like other publicly available datasets used for evaluation of DDoS \cite{sharafaldin2018toward} and IDS \cite{sharafaldin2018toward} detection methods, the datasets that we use here are highly imbalanced \cite{quiring2022and}. In CICDDoS2019, there are significantly more attack flow instances than benign flows, while in CICIDS2017 there are significantly more benign flow instances than attack flows.  Hence, to address the benign and attack class imbalance, we train and test our classifiers with equal numbers of attack and benign flows, which increases the confidence that the model is learning the correct feature representations from the patterns in the traffic flows.
In addition, CICIDS2017 suffers from attack vectors imbalance, i.e., the number of examples among the different attack vectors classes is highly divergent in the dataset. Specifically, we do not consider the infiltration and heartbleed attack vectors in CICIDS2017 because of their small portion (only 11 and 36 samples are present in the dataset respectively) in the overall dataset.

For each one of the datasets, we randomly sampled 80,000/79992 benign/attack flows. Then we split the data into train (50\%) and test (50\%) sets, with 20\% of the train set used for validation.
Table \ref{tab:datasplit} summarizes the data split that we use for each dataset. In these considerations we follow similar balancing and sampling schemes used by other works that evaluate their binary classification models using CICDDoS2019 \cite{CIC2_ortet2021towards, CIC3_elsayed2020ddosnet, CIC5_novaes2020long, CIC8_almiani2021ddos} and CICIDS2017 \cite{roopak2019deep, doriguzzi2020lucid, belarbi2022intrusion}. This way we ensure proper baselines when comparing the performance and accuracy results of our method to these state-of-the-art classification models.

\begin{table}[t]
\small
\center
\caption{\label{table:dataset-split} Flow distribution among all datasets.}
\begin{tabular}{ p{1.5cm}  p{4.8cm}  r }
\hline
  Dataset & Flow Type & Quantity \\
\hline
\multirow{2}{*}{Training} & Benign & 32,000 \\
    & Attack & 31,996 \\
    \hline
\multirow{2}{*}{Validation} & Benign & 8,000 \\
    & Attack & 8,000 \\
    \hline
\multirow{2}{*}{Test} & Benign & 40,000 \\
    & Attack & 39,996 \\
\hline 
\end{tabular}
\label{tab:datasplit}
\end{table}


The Set-Tree model can be applied to a single decision tree, or using a gradient boosting algorithm applied to Set-Trees. Here, we choose the latter, by training our model over an ensemble of 10 trees. We use the default training parameters: each tree has a 10 depth limit, the attention-set limit is 5, i.e., the attention set is either the entire input set, or a subset of packets that was generated in the last 5 nodes leading to the current node.
The subset of $(\alpha, \beta)$ pairs is limited to correspond to the following group of operators $Ops=$ $\{min,$ $max,$ $sum,$ $mean,$ $harmonic$ $mean,$ $geometric$ $mean,$ $second$   $moment$  $mean$$\}$.

\section{Results} \label{sec:results}
In this section, we present the results obtained by our proposed detection method with the datasets presented in Section \ref{subsec:dataset}. 
Evaluation metrics (defined in the appendix for completeness) of Accuracy (ACC), Precision, Recall and F-Score are used for performance measurement and for comparison with state-of-the-art models.

We start with the detection performance over the CICDDoS2019 and CICIDS2017 datasets, and continue by comparing it to existing state-of-the-art techniques. Then we present the early-detection results observed by considering only the first packets of each input flow, using the \textit{same} classifier on which the former results were reported. 
We also provide a packet-based features importance list, obtained from our tree models. We then discuss the running time and the overhead of the proposed model compared to vanilla decision trees, as well as optimization methods. 

\subsection{Detection Performance} \label{subsec:datasets_detection_performance}
\subsubsection{CICDDoS2019} 
Table \ref{table:cicddos2019-results} presents the performance results of our proposed model over the CICDDoS2019 dataset. As described in Section \ref{subsec:training}, we use a single classifier that was trained using a 32,000/31,996 benign/attack samples including all the CICDDoS2019 attack vectors together. The evaluation is provided over the whole testing set, and on each attack's testing set alone, using this single classifier. As seen in Table \ref{table:cicddos2019-results}, our model accuracy reaches an accuracy of 0.999. 
Note that each row in the table represents a balanced testing set (i.e., equal numbers of benign and attack samples).

\pgfplotstableset{
  fixed five/.style={
    precision=5,
    fixed,
    fixed zerofill,
  },
}

\begin{table}[t]
    \centering
  \caption{Our proposed model detection performance
 using the CICDDoS2019 dataset}
  \pgfplotstabletypeset[
    col sep=comma,
    columns={Attack Test Set, Recall, Precision, F1 Score, Accuracy},
    columns/Attack Test Set/.style={string type,
      column type={p{2.3cm}},},
    every head row/.style={before row=\hline,after row=\hline},
    every last row/.style={before row=\hline\bfseries,after row=\hline},
    columns/Recall/.style={fixed five, column type={>{\raggedright\arraybackslash\small}p{1.1cm}}},
    columns/Precision/.style={fixed five, column type={>{\raggedright\arraybackslash\small}p{1.1cm}}},
    columns/Accuracy/.style={fixed five, column type={>{\raggedright\arraybackslash\small}p{1.1cm}}},
    columns/F1 Score/.style={fixed five, column type={>{\raggedright\arraybackslash\small}p{1.1cm}}},
    ]{
      Attack Test Set, Recall, Precision, F1 Score, Accuracy
      DNS,  1,	1,	1,	1
      LDAP, 0.99971, 1.0, 0.99985, 0.99985
      MSSQL, 1,	0.999451153,	0.999725501,	0.999725426
      NTP, 0.998074807,	1,	0.999036476,	0.999037404
      NetBIOS, 1,	1,	1,	1
      SNMP,1,	1,	1,	1
      SSDP,0.999724972,	1,	0.999862467,	0.999862486
      UDP,1,	0.999500375,	0.999750125,	0.999750062
      SYN, 0.999823944,	1,	0.999911964,	0.999911972
      TFTP, 1,	0.999725048,	0.999862505,	0.999862486
      UDP-Lag, 1,	0.999177632,	0.999588647,	0.999588477
      Overall, 0.999749975,	0.99984997,	0.99979997,	0.99979998
    }
  \label{table:cicddos2019-results}
\end{table}

\subsubsection{CICIDS2017} 
The binary detection performance of our Set-Tree classifier, trained on the CICIDS2017 dataset, is presented in Table \ref{table:cicids2017-results}. Although this dataset contains a variety of attack vectors that differ from the temporal dynamics of DDoS attacks, the results demonstrate that a single Set-Tree classifier can effectively distinguish between patterns of benign and attack flows, achieving an overall accuracy of over 0.996. Note that two attack vectors, namely the {\textit Bot attack} and the {\textit Web-attack Brute-Force}, exhibit relatively lower detection accuracy values of 0.916 and 0.934, respectively. Here, the Bot and Web-attack Brute-Force vectors demonstrate detection with high precision (i.e., 1 and 0.99 respectively) but with lower recall values (i.e., 0.831 and 0.877 respectively). This indicates that the lower accuracy is due to instances where the classifier falsely identifies the corresponding attack flows as benign flows.

A bot attack involves the use of automated scripts to engage in malicious actions such as data theft, fraudulent purchases, or impersonating legitimate visitors. In the Brute-force web attack, the attacker crafts and sends web requests to a server using predefined values, in order to analyze the server's response. In the case of the CICIDS2017 dataset, this attack vector includes authentication attack flows in which the attacker tries a list of passwords to find the administrator’s password. The nature of these two attack vectors may explain their relatively lower recall and their resemblance to benign behavior.

\begin{table}[t]
    \centering
  \caption{Our proposed model detection performance
 using the CICIDS2017 dataset.}
  \pgfplotstabletypeset[
    col sep=comma,
    columns={Attack Test Set, Recall, Precision, F1 Score, Accuracy},
    columns/Attack Test Set/.style={string type,
      column type={p{2.3cm}},},
    every head row/.style={before row=\hline,after row=\hline},
    every last row/.style={before row=\hline\bfseries,after row=\hline},
    columns/Recall/.style={fixed five, column type={>{\raggedright\arraybackslash\small}p{1.1cm}}},
    columns/Precision/.style={fixed five, column type={>{\raggedright\arraybackslash\small}p{1.1cm}}},
    columns/Accuracy/.style={fixed five, column type={>{\raggedright\arraybackslash\small}p{1.1cm}}},
    columns/F1 Score/.style={fixed five, column type={>{\raggedright\arraybackslash\small}p{1.1cm}}},
    ]{
      Attack Test Set, Recall, Precision, F1 Score, Accuracy
      GoldenEye,  0.9930939226519337,0.9986111111111111,0.995844875 ,0.9958563535911602
      DDoS, 0.9981775300171527,0.9957223826328735,0.996948445 ,0.9969446826758147
      Bot, 0.8311688311688312,1.0, 0.907801418 ,0.9155844155844156
      DoS Hulk, 0.9977976263306008,0.995665710274098, 0.996730528,0.9967270280190872
      Slowhttptest, 1.0,0.9950617283950617, 0.997524752,0.9975186104218362
      DoS slowloris,1, 1,	1,	1
      FTP-Patator,0.9963636363636363,0.9945553539019963,0.995458674 ,0.9954545454545455
      PortScan,0.9996515679442509,0.9960940890547696,0.997869658 ,0.9978658536585366
      SSH-Patator, 0.9844789356984479,0.9932885906040269,0.988864143 , 0.9889135254988913
      WA Brute Force, 0.8771929824561403,0.9900990099009901,0.930232558 ,0.9342105263157895
      WA XSS, 0.9761904761904762,0.9761904761904762,0.976190476 ,0.9761904761904762
      Overall, 0.9971995799369906,0.995830108117556,0.996514374,0.9965119767965195
    }
  \label{table:cicids2017-results}
\end{table}

\subsection{State-of-The-Art Comparison}  \label{subsec:results-sota}
\subsubsection{CICDDoS2019} 
We compare the performance results of our model to existing state-of-the-art models that evaluated their work on the CICDDoS2019 dataset. To distinguish between our approaches, we denote our classifier as ST-full when tested with complete flow streams as inputs. Additionally, for the early detection variants, we use the notation ST-$p$, where $p$ represents the number of packets in the input stream prefix. 
Table \ref{table:sota_DDoS2019} and Figure \ref{figure:sota_DDoS2019} show that our set-tree model matches or outperforms other state-of-the-art models in all the performance metrics. The first row in the table (ST-full), presents our model performance with complete flow streams as input, while the second row in the table (ST-2),  presents the early detection results using the same classifier but with the first two packets for each flow as input. As seen in the first and second rows, our proposed classifier exhibits the same performance results for early-detection and detection of complete flows in CICDDoS2019. The detailed early-detection results, as well as the time-saving evaluation are further discussed in Section \ref{subsec:early-detection-DDoS2019}. 

\begin{table}[t]
    \centering
  \caption{Performance comparison against state-of-the-art methods using the CICDDoS2019 dataset.}
  \label{table:sota_DDoS2019}
  \pgfplotstabletypeset[
    col sep=comma,
    columns={Study, Method, Recall, Precision, F1 Score, Accuracy},
    columns/Method/.style={
      string type,
      column type={>{\raggedright\arraybackslash\small}p{1.2cm}},
    },
    columns/Study/.style={string type,
      column type={>{\raggedright\arraybackslash\small}p{1.9cm}},
      },
    every head row/.style={before row=\hline,after row=\hline},
    every last row/.style={after row=\hline},
    columns/Recall/.style={fixed five, column type={>{\raggedright\arraybackslash\small}p{0.8cm}}},
    columns/Precision/.style={fixed five, column type={>{\raggedright\arraybackslash\small}p{0.8cm}}},
    columns/Accuracy/.style={fixed five, column type={>{\raggedright\arraybackslash\small}p{0.9cm}}},
    columns/F1 Score/.style={fixed five, column type={>{\raggedright\arraybackslash\small}p{0.9cm}}},
    ]{
      Study, Method, Recall, Precision, F1 Score, Accuracy
      Our Study, ST-full,  0.999749975,	0.99984997,	0.99979997,	0.99979998
      Our Study,ST-2, 0.99975,0.99975,0.99975,0.99975
      CyDDoS\cite{CIC2_ortet2021towards},DNN, 0.99650, 0.99650,0.99600, 0.99620
      DDoSNet\cite{CIC3_elsayed2020ddosnet},RNN+AE, 0.99000, 0.99000, 0.99000, 0.98800 
      FUZZY\cite{CIC5_novaes2020long}, LSTM ,0.93130, 0.97890,0.95450, 0.96220 
      Kalman\cite{CIC8_almiani2021ddos}, DNN,0.97490, 0.91220,0.94300, 0.94000 
    }
  
\end{table}

CyDDoS \cite{CIC2_ortet2021towards} and DDoSNet  
\cite{CIC3_elsayed2020ddosnet} perform almost as well as our model. However, these approaches (as well as the works compared in Figure \ref{figure:sota_DDoS2019}) employ in their detection deep learning techniques such as RNN and CNN, which are computationally demanding.

Note that for a fair baseline, as discussed earlier in Section \ref{subsec:training}, in our work we follow similar evaluation methodologies as reported by other state-of-the-art works compared here. For example, as reported by \cite{CIC2_ortet2021towards}, their evaluation is performed only on a balanced set of 52,231/51,746 benign/attack sampled flows out of the CICDDoS2019 dataset. Similarly, in \cite{CIC3_elsayed2020ddosnet}, the authors evaluated their proposed model on a balanced subset of 23,000 sampled flows.

\definecolor{color1}{RGB}{54,197,239}
\definecolor{color2}{RGB}{223,30,90}
\definecolor{color3}{RGB}{45,182,124}
\definecolor{color4}{RGB}{235,177,45}
\definecolor{color5}{RGB}{74,21,75}
\definecolor{color10}{RGB}{44,160,44}
\begin{figure}[t]
  \centering
  \begin{tikzpicture}
    \begin{axis}[
      ybar,
      ymin=0.85,
      ymax=1.03,
      width=0.53\textwidth,
      height=6cm,
      symbolic x coords={Recall, Precision, Accuracy, F1 Score},
      ytick={0.85,0.86,0.87,0.88,0.89,0.90,0.91,0.92,0.93,0.94,0.95,0.96,0.97,0.98,0.99,1.0},
      tick label style={font=\footnotesize}, 
      xtick=data,
      nodes near coords={
        \pgfmathprintnumber[precision=4]{\pgfplotspointmeta}
      },
      grid=major,
      nodes near coords align={vertical},
      nodes near coords style={font=\tiny, rotate=90, anchor=west},
      bar width=5pt, 
      legend style={
        at={(0.43,-0.18)},legend columns=-1,font=\scriptsize,
        anchor=north, cells={align=left},
      },
    ]
      \addplot[fill=color1!90] coordinates {(Recall, 0.999749975) (Precision, 0.99984997) (Accuracy, 0.99979998) (F1 Score, 0.99979997)};
      \addplot[fill=color2!90] coordinates {(Recall, 0.99975) (Precision, 0.99975) (Accuracy, 0.99975) (F1 Score, 0.99975)};
      \addplot[fill=color3!90] coordinates {(Recall, 0.99650) (Precision, 0.99650) (Accuracy, 0.99620) (F1 Score, 0.99600)};
      \addplot[fill=color4!90] coordinates {(Recall, 0.99000) (Precision, 0.99000) (Accuracy, 0.98800) (F1 Score, 0.99000)};
      \addplot[fill=color5!90] coordinates {(Recall, 0.93130) (Precision, 0.97890) (Accuracy, 0.96220) (F1 Score, 0.95450)};
      \addplot[fill=purple!30] coordinates {(Recall, 0.97490) (Precision, 0.91220) (Accuracy, 0.94000) (F1 Score, 0.94300)};
      
      \legend{
        {ST-full},
        {ST-2.},
        {CyDDoS}, {DDoSNet},
        {FUZZY},
        {Kalman}
      }
    \end{axis}
  \end{tikzpicture}
  \caption{Performance comparison against state-of-the-art methods \cite{CIC2_ortet2021towards}, \cite{CIC3_elsayed2020ddosnet}, \cite{CIC5_novaes2020long}, \cite{CIC8_almiani2021ddos} using the CICDDoS2019 dataset. ST-full and ST-2 refer to our proposed model, evaluated on complete flow-stream inputs, and on first 2-packets inputs respectively.}
  \label{figure:sota_DDoS2019}
\end{figure}
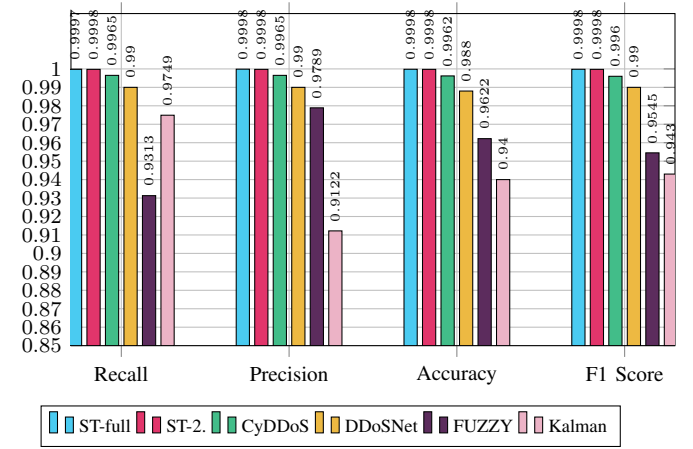

\subsubsection{CICIDS2017} 
\begin{figure}[t]
  \centering
  \begin{tikzpicture}
    \begin{axis}[
      ybar,
      ymin=0.85,
      ymax=1.03,
      width=0.53\textwidth,
      height=6cm,
      symbolic x coords={Recall, Precision, Accuracy, F1 Score},
      ytick={0.85,0.86,0.87,0.88,0.89,0.9,0.91,0.92,0.93,0.94,0.95,0.96,0.97,0.98,0.99,1.0},
      tick label style={font=\footnotesize}, 
      xtick=data,
      nodes near coords={
        \pgfmathprintnumber[precision=4]{\pgfplotspointmeta}
      },
      grid=major,
      nodes near coords align={vertical},
      nodes near coords style={font=\tiny, rotate=90, anchor=west},
      bar width=5pt, 
      legend style={
        at={(0.46,-0.18)},legend columns=-1,font=\scriptsize,
        anchor=north, cells={align=left},
      },
    ]
      \addplot[fill=color1!90] coordinates {(Recall, 0.99720) (Precision, 0.99583) (Accuracy, 0.99651) (F1 Score, 0.99651)};
      \addplot[fill=color2!90] coordinates {(Recall, 0.94184) (Precision, 0.99291) (Accuracy, 0.96756) (F1 Score, 0.96670)};
      \addplot[fill=color4!90] coordinates {(Recall, 0.89890) (Precision, 0.98440) (Accuracy, 0.96240) (F1 Score, 0.93970)};
      \addplot[fill=color5!90] coordinates {(Recall, 0.90170) (Precision, 0.98140) (Accuracy, 0.95140) (F1 Score, 0.93986)};
      \addplot[fill=purple!30] coordinates {(Recall, 0.99100) (Precision, 0.97410) (Accuracy, 0.97160) (F1 Score, 0.98248)};
      \addplot[fill=color3!90] coordinates {(Recall, 0.98855) (Precision, 0.94475) (Accuracy, 0.96665) (F1 Score, 0.96615)};
      
      \legend{
        {ST-full},
        {ST-4.},
        {LSTM},
        {1D-CNN},
        {1D-CNN+LSTM},{DT}
      }
    \end{axis}
  \end{tikzpicture}
  \caption{Performance comparison against state-of-the-art methods \cite{roopak2019deep}, \cite{ahmim2019novel} using the CICIDS2017 dataset. ST-full and ST-4 refer to our proposed model evaluated on complete flow-stream inputs, and on first 4-packets inputs respectively.}
  \label{figure:sota_IDS2017}
\end{figure}
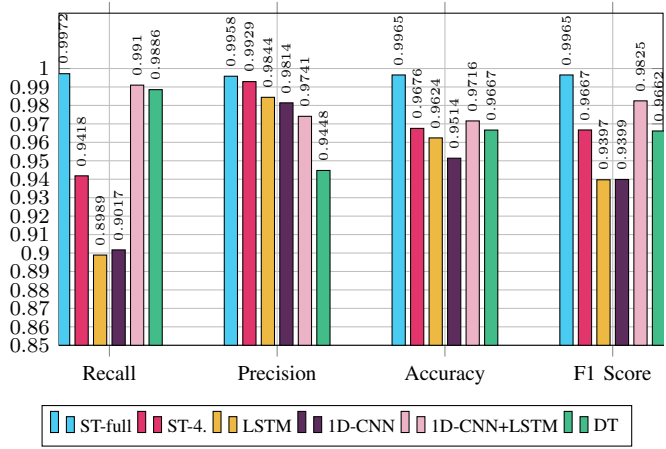

We also compare our set-tree model against the state-of-the-art intrusion detection approaches on the CICIDS2017 dataset, which has been widely used for evaluating intrusion detection methods. For a fair comparison, we focus on binary classification methods, rather than multi-class classification. We selected the top-performing binary classification approaches for comparison, all trained using supervised learning methods. Table \ref{table:sota_IDS2017} and Figure \ref{figure:sota_IDS2017} compare our set-tree classifier performance results to four state-of-the-art models.

\begin{table}[t]
    \centering
  \caption{Performance comparison against state-of-the-art methods using the CICIDS2017 dataset.}
  
  \pgfplotstabletypeset[
    col sep=comma,
    columns={Study, Method, Recall, Precision, F1 Score, Accuracy},
    columns/Method/.style={
      string type,
      column type={>{\raggedright\arraybackslash\small}p{1.5cm}},
    },
    columns/Study/.style={string type,
      column type={>{\raggedright\arraybackslash\small}p{1.4cm}},
      },
    every head row/.style={before row=\hline,after row=\hline},
    every last row/.style={after row=\hline},
    columns/Recall/.style={fixed five, column type={>{\raggedright\arraybackslash\small}p{0.8cm}}},
    columns/Precision/.style={fixed five, column type={>{\raggedright\arraybackslash\small}p{0.8cm}}},
    columns/Accuracy/.style={fixed five, column type={>{\raggedright\arraybackslash\small}p{1.0cm}}},
    columns/F1 Score/.style={fixed five, column type={>{\raggedright\arraybackslash\small}p{0.9cm}}},
    ]{
      Study, Method, Recall, Precision, F1 Score, Accuracy
      Our Study, ST-full, 0.9971995799369906,0.995830108117556,0.996514374,0.9965119767965195
      Our Study, ST-4, 0.9418412761914288,0.9929092969923821,0.966701312,0.9675576336450468
      \cite{roopak2019deep},LSTM, 0.8989, 0.9844, 0.93970, 0.9624
      \cite{roopak2019deep}, 1D-CNN ,0.9017, 0.9814, 0.93986, 0.9514
      \cite{roopak2019deep}, \scriptsize{CNN+LSTM},0.991, 0.9741, 0.98248, 0.9716
      \cite{ahmim2019novel}, Decision Trees, 0.98855, 0.94475,0.96615, 0.96665  
    }
    \label{table:sota_IDS2017}
\end{table}

\begin{table}[t]
    \centering
  \caption{Performance comparison of our work vs.\ LUCID \cite{doriguzzi2020lucid} using the DDoS attack trace in the CICIDS2017 dataset.}
  
  \pgfplotstabletypeset[
    col sep=comma,
    columns={Study, Method, Recall, Precision, F1 Score, Accuracy},
    columns/Method/.style={
      string type,
      column type={>{\raggedright\arraybackslash\small}p{1.1cm}},
    },
    columns/Study/.style={string type,
      column type={>{\raggedright\arraybackslash\small}p{1.6cm}},
      },
    every head row/.style={before row=\hline,after row=\hline},
    every last row/.style={after row=\hline},
    columns/Recall/.style={fixed five, column type={>{\raggedright\arraybackslash\small}p{0.8cm}}},
    columns/Precision/.style={fixed five, column type={>{\raggedright\arraybackslash\small}p{0.8cm}}},
    columns/Accuracy/.style={fixed five, column type={>{\raggedright\arraybackslash\small}p{1.0cm}}},
    columns/F1 Score/.style={fixed five, column type={>{\raggedright\arraybackslash\small}p{0.9cm}}},
    ]{
      Study, Method, Recall, Precision, F1 Score, Accuracy
      Our Study, ST-full, 0.998625,0.99650742,0.99757,0.99756
    LUCID\cite{doriguzzi2020lucid}, CNN,0.999, 0.993,0.9966, 0.9967
    }
    \label{table:sota_IDS2017_lucid}
\end{table}

The first row in the table (ST-full) refers to the performance results of our model considering the complete flows as input, while the second row in the table (ST-4) refers to the early-detection results considering the first four packets of each flow as input. Here, the early-detection results still introduce a significant time-saving, but exhibit relatively lower accuracy results than the complete-flow detection results, as we elaborate more in Section \ref{subsec:early-detection-IDS2017}. LSTM, 1D-CNN and 1D-CNN+LSTM \cite{roopak2019deep} (the latter one is a combination of the former ones) are deep learning models performing binary classification using the Internet of Things (IoT) networks \cite{roopak2019deep}. The input layer of these models includes 82 units, which correspond to each one of the aggregated flow features provided in the CICIDS2017 dataset, while the output layer returns a probability used to perform binary classification.

LUCID \cite{doriguzzi2020lucid} is another work that achieved high performance results using CNNs and evaluated their results on the CICIDS2017 dataset. However, they evaluated their work only on a single trace out of CICIDS2017, which represents the DDoS attack vector. In this work, we report the evaluation results over the entire dataset, including all its traces and attack vectors that represent various cyber-attacks. To have a fair comparison with LUCID \cite{doriguzzi2020lucid}, we trained  an additional classifier using our approach on their single attack vector. We followed the same training and testing methodology as described in Section \ref{subsec:training}, i.e., balanced training, validation and test sets, which matches the experiments reported in \cite{doriguzzi2020lucid}. A comparison of our classifier detection results vs. LUCID over the DDoS attack trace in CICIDS2017 is presented in Table \ref{table:sota_IDS2017_lucid}.

\subsection{Early Detection} \label{subsec:early-detection-results}
Here, our goal is to study how early our model can distinguish between benign and attack flows with high accuracy.
Since our detection model operates on a stream-of-packets as its flow input, we address this question by testing the same classifier that was trained on the complete flow streams, with various $p$-sized-length sub-streams of packets, where each sub-stream contains the first $p$ packets of the flow.

\begin{table*}[t]
\caption{\label{table:ddos-early-detection-results-2017} Early Detection Performance Results for CICIDS2017.}
\footnotesize
\center
\begin{tabular}{ m{1.8cm}|m{0.9cm} m{1.1cm} l|m{0.9cm} m{1.1cm} l|m{0.9cm} m{1.1cm} l| m{0.9cm}  l }
\hline
  & \multicolumn{3}{c|}{ST-2 (First 2-Packets)} & \multicolumn{3}{c|}{ST-4} & \multicolumn{3}{c|}{ST-14} & \multicolumn{2}{c}{ST-full \scriptsize{(Complete Flow)}} \\
\hline
Vector & Mean \newline Duration (ms)& \textbf{Time} \newline \textbf{Saving \% } & Accuracy & Mean \newline Duration (ms)& \textbf{Time} \newline \textbf{Saving \% }  & Accuracy & Mean \newline Duration (ms)& \textbf{Time} \newline \textbf{Saving \% } & Accuracy & Mean \newline Duration (ms) & Accuracy \\
\hline
\scriptsize{DoS GoldenEye} & \multicolumn{1}{r}{11900} & \begingroup \color{blue} {\tiny$\blacktriangledown$}38.6\% \endgroup & 0.9385 
& \multicolumn{1}{r}{11968} & \begingroup \color{blue} {\tiny$\blacktriangledown$}38.3\% \endgroup & 0.9571
& \multicolumn{1}{r}{18786} & \begingroup \color{blue} {\tiny$\blacktriangledown$}3.2\% \endgroup & 0.9568
& \multicolumn{1}{r}{19408} 
& {0.9959}\\
\hline
DoS slowloris & \multicolumn{1}{r}{470} & \begingroup \color{blue} {\tiny$\blacktriangledown$}99.17\% \endgroup & 0.9515 
& \multicolumn{1}{r}{22320} & \begingroup \color{blue} {\tiny$\blacktriangledown$}60.5\% \endgroup & 0.9440
& \multicolumn{1}{r}{34949} & \begingroup \color{blue} {\tiny$\blacktriangledown$}38.2\% \endgroup & 0.9731
& \multicolumn{1}{r}{56554} 
& 1.0000\\
\hline
DoS Hulk & \multicolumn{1}{r}{1521} & \begingroup \color{blue} {\tiny$\blacktriangledown$}97.28\% \endgroup & 0.9075
& \multicolumn{1}{r}{16895} & \begingroup \color{blue} {\tiny$\blacktriangledown$}69.9\% \endgroup & 0.9818
& \multicolumn{1}{r}{55599} & \begingroup \color{blue} {\tiny$\blacktriangledown$}0.79\% \endgroup & 0.9951
& \multicolumn{1}{r}{56044}
& 0.9967\\
\hline
{Slowhttptest} & \multicolumn{1}{r}{1131} & \begingroup \color{blue} {\tiny$\blacktriangledown$}98.07\% \endgroup & 0.9243
& \multicolumn{1}{r}{5602} & \begingroup \color{blue} {\tiny$\blacktriangledown$}90.44\% \endgroup & 0.9615
& \multicolumn{1}{r}{55492} & \begingroup \color{blue} {\tiny$\blacktriangledown$}5.32\% \endgroup & 0.9839
& \multicolumn{1}{r}{58615}
& 0.9975\\
\hline
{Brute Force} & \multicolumn{1}{r}{0.116} & \begingroup \color{blue} {\tiny$\blacktriangledown$}99.99\% \endgroup & 0.8860
& \multicolumn{1}{r}{4724} & \begingroup \color{blue} {\tiny$\blacktriangledown$}27.32\% \endgroup & {0.9342}
& \multicolumn{1}{r}{5033} & \begingroup \color{blue} {\tiny$\blacktriangledown$}22.58\% \endgroup & {0.9342}
& \multicolumn{1}{r}{6502}
& 0.9342 \\
\hline
WA XSS & \multicolumn{1}{r}{0.128} & \begingroup \color{blue} {\tiny$\blacktriangledown$}99.99\% \endgroup & 0.9405
& \multicolumn{1}{r}{5101} & \begingroup \color{blue} {\tiny$\blacktriangledown$}22.84\% \endgroup & 0.9643
& \multicolumn{1}{r}{5184} & \begingroup \color{blue} {\tiny$\blacktriangledown$}21.58\% \endgroup & 0.9762
& \multicolumn{1}{r}{6611}
& 0.9762 \\
\hline
PortScan & \multicolumn{1}{r}{0.060} & \begingroup \color{blue} {\tiny$\blacktriangledown$}99.93\% \endgroup & 0.9460
& \multicolumn{1}{r}{80} & \begingroup \color{blue} {\tiny$\blacktriangledown$}3.18\% \endgroup & 0.9958
& \multicolumn{1}{r}{81} & \begingroup \color{blue} {\tiny$\blacktriangledown$}1.56\% \endgroup & 0.9978
& \multicolumn{1}{r}{82}
& 0.9979 \\
\hline
Bot & \multicolumn{1}{r}{0.503} & \begingroup \color{blue} {\tiny$\blacktriangledown$}99.86\% \endgroup & {\color{red} 0.7110}
& \multicolumn{1}{r}{126} & \begingroup \color{blue} {\tiny$\blacktriangledown$}63.91\% \endgroup & 0.9156
& \multicolumn{1}{r}{328} & \begingroup \color{blue} {\tiny$\blacktriangledown$}6.51\% \endgroup & 0.9156
& \multicolumn{1}{r}{350}
& 0.9156 \\
\hline
DDoS & \multicolumn{1}{r}{4.8} & \begingroup \color{blue} {\tiny$\blacktriangledown$}99.97\% \endgroup & {\color{red} 0.4569}
& \multicolumn{1}{r}{1099} & \begingroup \color{blue} {\tiny$\blacktriangledown$}93.05\% \endgroup & 0.9327
& \multicolumn{1}{r}{{\scriptsize15352}} & \begingroup \color{blue} {\tiny$\blacktriangledown$}3.07\% \endgroup & 0.9952
& \multicolumn{1}{r}{15837}
& 0.9969 \\
\hline
FTP-Patator & \multicolumn{1}{r}{14.7} & \begingroup \color{blue} {\tiny$\blacktriangledown$}99.6\% \endgroup & {\color{red} 0.7164} 
& \multicolumn{1}{r}{19} & \begingroup \color{blue} {\tiny$\blacktriangledown$}99.5\% \endgroup & {\color{red} 0.7382}
& \multicolumn{1}{r}{1540} & \begingroup \color{blue} {\tiny$\blacktriangledown$}65.8\% \endgroup & 0.9936
& \multicolumn{1}{r}{4513} 
& 0.9955\\
\hline
SSH-Patator & \multicolumn{1}{r}{0.752} & \begingroup \color{blue} {\tiny$\blacktriangledown$}99.99\% \endgroup & {\color{red} 0.7838} 
& \multicolumn{1}{r}{2.5} & \begingroup \color{blue} {\tiny$\blacktriangledown$}99.96\% \endgroup & {\color{red} 0.7927}
& \multicolumn{1}{r}{62} & \begingroup \color{blue} {\tiny$\blacktriangledown$}99.01\% \endgroup & 0.9268
& \multicolumn{1}{r}{6168} 
& 0.9889\\
\hline
\hline
Overall & \multicolumn{1}{r}{547} & \begingroup \color{blue} {\tiny$\blacktriangledown$}97.16\% \endgroup & 0.8098
& \multicolumn{1}{r}{4758} & \begingroup \color{blue} {\tiny$\blacktriangledown$}75.31\% \endgroup & 0.9676
& \multicolumn{1}{r}{15465} & \begingroup \color{blue} {\tiny$\blacktriangledown$}19.77\% \endgroup &  0.9925
& \multicolumn{1}{r}{19277} 
& 0.9965\\
\hline
\label{table:early_results_ids2017}
\end{tabular}
\end{table*}

\begin{table}[t]
\caption{\label{table:ddos-early-detection-results-2019} Early Detection Performance Results for CICDDoS2019.}
\center
\footnotesize
\begin{tabular}{ m{1.1cm}| m{0.9cm} m{1.1cm} l| m{0.9cm}  l }
\hline
  & \multicolumn{3}{c|}{ST-2 \scriptsize{(First 2-Packets)}} & \multicolumn{2}{c}{ST-full \scriptsize{(Complete Flow)}} \\
\hline
Vector & Mean \newline Duration (ms)& \textbf{Time} \newline \textbf{Saving \% } & Accuracy & Mean \newline Duration (ms) & Accuracy \\
\hline
NTP & \multicolumn{1}{r}{0.494} &  \begingroup
    \color{blue}
    {\tiny$\blacktriangledown$}96.8\%
  \endgroup & 0.99917 & \multicolumn{1}{r}{15.407} & 0.99904\\
\hline
TFTP & \multicolumn{1}{r}{0.062} &  \begingroup
    \color{blue}
    {\tiny$\blacktriangledown$}99.99\%
  \endgroup & 0.99986 & \multicolumn{1}{r}{2010} & 0.99986 \\
\hline
SNMP & \multicolumn{1}{r}{0.006} & \begingroup
    \color{blue}
    {\tiny$\blacktriangledown$}99.5\%
  \endgroup & 0.99972 & \multicolumn{1}{r}{1.280} & 1.00000 \\
\hline
LDAP & \multicolumn{1}{r}{0.007} & \begingroup
    \color{blue}
    {\tiny$\blacktriangledown$}98.5\%
  \endgroup & 0.99986 & \multicolumn{1}{r}{0.467} & 0.99985 \\
\hline
NetBIOS & \multicolumn{1}{r}{0.010} & \begingroup
    \color{blue}
    {\tiny$\blacktriangledown$}99.97\%
  \endgroup & 1.00000 & \multicolumn{1}{r}{28.692} & 1.00000 \\
\hline
MSSQL & \multicolumn{1}{r}{0.006} & \begingroup
    \color{blue}
    {\tiny$\blacktriangledown$}99.9\%
  \endgroup& 0.99973 & \multicolumn{1}{r}{6.374} & 0.99973\\
\hline
SYN & \multicolumn{1}{r}{0.131} & \begingroup
    \color{blue}
    {\tiny$\blacktriangledown$}99.99\%
  \endgroup & 0.99982 & \multicolumn{1}{r}{8088} & 0.99991 \\
\hline
UDP  & \multicolumn{1}{r}{0.011} & \begingroup
    \color{blue}
    {\tiny$\blacktriangledown$}99.99\%
  \endgroup & 0.99975 & \multicolumn{1}{r}{89.865} & 0.99975 \\
\hline
SSDP & \multicolumn{1}{r}{0.008} & \begingroup
    \color{blue}
    {\tiny$\blacktriangledown$}99.99\%
  \endgroup & 0.99986 & \multicolumn{1}{r}{82.636} & 0.99986 \\
\hline
\scriptsize{UDP-LAG} & \multicolumn{1}{r}{0.049} & \begingroup
    \color{blue}
    {\tiny$\blacktriangledown$}99.99\%
  \endgroup & 0.99959 & \multicolumn{1}{r}{5222} & 0.99959 \\
\hline
DNS & \multicolumn{1}{r}{0.068} & \begingroup
    \color{blue}
    {\tiny$\blacktriangledown$}94.35\%
  \endgroup & 0.99972 & \multicolumn{1}{r}{1.203} & 1.00000 \\
\hline 
\hline 
Overall & \multicolumn{1}{r}{10.391} & \begingroup
    \color{blue}
    {\tiny$\blacktriangledown$}99.79\%
  \endgroup & 0.99975 & \multicolumn{1}{r}{4918.738} & 0.99980 \\
\hline 
\end{tabular}
\label{table:early_results_ddos2019}
\end{table}

The decision to use the same classifier trained on complete flows for evaluating early detection is a critical aspect that highlights the advantages of our detection method. The set-compatible-split-criterion and the corresponding attention mechanism of the set-tree model enable decision trees to operate on subsets of packets (as described earlier in Section \ref{subsec:set-tree-model}). This may become powerful in learning sub-patterns that reside in the first few packets of a flow.
While developing separate classifiers for early detection, that are trained using data from the first $p$ packets in each flow might yield higher performance, it would compromise the flexibility and simplicity of our proposed solution. Managing multiple classifiers based on sub-flow packet counts could lead to complexity and impracticality. Additionally, relying on a single classifier that can detect variable-length sub-streams of packets, in which the detection accuracy is correlated to the number of first packets received as input, opens the door for a variety of dynamic detection strategies aligned with system requirements and resource constraints.

Given the flows packet count and duration analysis provided in Section \ref{subsec:dataset} for both CICDDoS2019 and CICIDS2017 datasets, the mean duration values of the first two packets sub-flows are lower by several orders of magnitude than the complete flows mean duration. In CICDDoS2019, the mean duration of the first two packets sub-flows is 10 ms compared to 4.91 seconds for the complete flows, while in CICIDS2017, it is 546 ms compared to 19.27 seconds. Based on this observation, we start our early-detection experiments by using the first two packets as input flows (ST-2), and gradually increase the input packet count until the performance of the complete flows is observed.

To enable this experimentation, we utilized packet-based data-processing, as described in Section \ref{subsec:dataprocess}, which involved reconstructing the flows into stream-of-packets structures while preserving the original timestamps of each packet. This approach allows us to extract and experiment with the first packet sub-streams as input samples from each dataset.

\subsubsection{CICDDoS2019} \label{subsec:early-detection-DDoS2019}
Table \ref{table:early_results_ddos2019} presents the early-detection results in CICDDoS2019.
Here, the early-detection accuracy of the first-two packets input (ST-2), is compared with the detection accuracy results of the complete-flows (ST-full). As in Section \ref{subsec:datasets_detection_performance}, the early-detection evaluation is provided over the whole testing set, and on each attack's testing set alone.
Note that here as well, each one of the rows in the table represents a balanced testing set (i.e., equal numbers of benign and attack samples).

The second column in the table contains the mean duration value of the first two packets' flows (ST-2), as well as the time-saving in percentages (TS\%). We define TS\% as: 
\[TS\%=\{100\times(1 - \frac{d} {D} )\},\]
where $d$ is the mean duration of the first two packets, and $D$ is the mean duration of the complete flows. As seen in Table \ref{table:early_results_ddos2019}, the ST-2 early-detection results (using first two packets) match the complete flows detection, with a mean time-saving of 99.79\%. This high performance is maintained across all the attack vectors in CICDDoS2019, indicating that the first two packets of each flow are sufficient to distinguish between benign and DDoS traffic in this dataset. The recall, precision and F-1 results that correspond to the overall CICDDoS2019 test-set were provided in Table \ref{table:sota_DDoS2019}.

\subsubsection{CICIDS2017}  \label{subsec:early-detection-IDS2017}
The early-detection results for the CICIDS2017 dataset are presented in Table \ref{table:early_results_ids2017}.
Similar to the early-detection results in CICDDoS2019, we compare between detection accuracy results of first-packets sub-streams (ST-x) and the complete flow streams (ST-full). However, here the detection accuracy (0.8098) over the first two packets' flow inputs (ST-2) is significantly lower than the complete flow (ST-full) detection accuracy (0.9965). Thus, we gradually increase the first packets count of the flow inputs, and provide the performance results for four-packets (ST-4), and fourteen-packets (ST-14). Here as well, for each input size we provide time-saving in percentages (TS\%).

The results show that while the early-detection performance based on the first two packets (ST-2) is high for some attack vectors (e.g., the PortScan attack vector has a detection accuracy of 0.9460 with a time-saving of 99.99\%), few attack vectors such as the Bot, DDoS, FTP-Patator and SSH-Patator have low detection accuracy (0.7110, 0.4569, 0.7164 and 0.7838 respectively). The ST-4 overall detection accuracy based on the first four-packets is high (0.9676), exhibiting a mean time-saving of 75.31\% (mean duration of 4.76 seconds for ST-4 compared with a mean duration of 19.28 seconds for ST-full on complete-flows). However, while most of the attack vectors' detection accuracy based on the first four-packets matches the complete-flows detection accuracy, two attack vectors, namely the FTP-Patator and the SSH-Patator detection accuracy is still far from the corresponding complete-flows detection accuracy.  These attack vectors have the highest packet-count median values among the attack vectors in CICIDS2017 (12 and 26 respectively, as appears in Table \ref{tab:ids_duration_table} in Section \ref{sec:method}.

As seen in Table \ref{table:early_results_ids2017}, for each one of the attack vectors in CICIDS2017, the proposed early-detection method achieves high accuracy on sub-streams of its first-packets which exhibit a significant lower duration than the complete flow duration.




\subsection{Features Importance Analysis} \label{subsec:features}
The Set-Tree model supports an extraction of feature importance values from the trained model. The feature importance is calculated by quantifying the total reduction (normalized) in the criteria achieved by that specific feature. This metric is commonly referred to as the \textit{Gini} importance. Since our model is composed of gradient-boosted Set-Trees, the feature importance values are computed as the mean importance value across all the trees in the model. 

\begin{table}[t]
\caption{\label{table:features} Feature importance values extracted from our models using CICDDoS2019 and CICIDS2017}
\small
\center
    \begin{tabular}{m{2.3cm} m{1.2cm} m{2.3cm} m{1.1cm}}
    \hline
      \multicolumn{2}{c}{Set Tree - CICDDoS2019} & \multicolumn{2}{c}{Set Tree - CICIDS2017}\\
    \cline{1-4}
     Feature & Importance & Feature & Importance \\ \hline
         length & 0.76667 & is\_RST\_flag & 0.44361 \\ 
         is\_SYN\_flags & 0.22735 & length & 0.25440 \\ 
         direction\_iat\_before & 0.00256 & is\_forward & 0.06298 \\ 
         direction\_iat\_after & 0.00216 & is\_ACK\_flag & 0.03615 \\ 
         iat\_before & 0.00092 & init\_win\_bytes & 0.03359 \\ 
         iat\_after & 0.00030 & iat\_after & 0.01957 \\ \hline
    \end{tabular}
\end{table}

Table \ref{table:features} provides a comparison between the high-importance features of the CICDDoS2019 set-tree classifier and the CICIDS2017 set-tree classifier.
Note that the features in our detection model represent packet header features rather than aggregated flow features. Due to the nature of the set-compatible split criterion of the Set-Tree model (as described in Section \ref{subsec:set-tree-model}), a packet feature may represent different statistical functions (minimum, maximum, sum, mean etc) that are used in the Set-Trees' nodes as split criteria (e.g., the packet's length feature importance may represent various functions that are used in our model, such as sum of all packets' length, min packet length etc).       

The table shows us that the packet length feature is a dominant feature with high importance across both classifiers. 


\subsection{Time Complexity} \label{subsec:runtime}
Tree-learning algorithms implement exhaustive search for the optimal decision nodes’ parameters. The set-compatible criterion, introduced by Set-Tree, adds complexity to the \textit{learning} process. As described by \cite{pmlr-v139-hirsch21a}, the overhead learning factor is $2N_{\alpha,\beta}(N_h+1)$, where $N_{\alpha,\beta}$ is the number of statistical operations that are represented by the $\alpha$, $\beta$ pair, and $N_h$ is the number of nodes scanned for attention sets along a decision path. This overhead can be further reduced by caching mechanisms and sampling techniques. As for the running time, tree-based algorithms are considered extremely fast compared to other complex methods, e.g., neural networks. The Set-Tree implementation that we used in our work (as provided by \cite{pmlr-v139-hirsch21a}) has a computational overhead due to its sub-optimal implementation. Note that Set-Tree differs from a vanilla decision tree only in the decision rules. Thus, its implementation can benefit from almost any existing optimization for tree-learning algorithms.



\section{Conclusions} \label{sec:discussion}
In this work, we have presented a new approach to intrusion detection that considers a flow input as a stream of packet headers instead of using traditional aggregated flow records. 
We demonstrated how a tree-based classifier that operates on the proposed flow input matches or exceeds existing state-of-the-art machine learning techniques, including Deep Learning methods on two recent and comprehensive datasets. 
Furthermore, we show that the proposed stream input enables early detection with a significant time-saving, by relying on the first few packets of the flow. The fundamental benefit of early detection using our approach is twofold. First, a stream of packet headers is a natural data structure for multiple-time-window detection. Second, as we empirically show, the same classifier that is trained on the complete streams, is able at detection time to infer malicious sub-streams, composed of the first few packets of the complete stream - usually 2 or 4 packets suffice for highly accurate detection.   Note that we did not re-train the model for the small packet size. Moreover, our tree-based model offers better interpretability in comparison to deep learning and other complex models.

The idea of developing a fast and early detection mechanism that relies on a fraction of the traffic is appealing. As a response to the fast growth in size and frequency of DDoS attacks, detection time becomes a key factor in reducing the potential damage and disruption to the targeted network.  
As we show on multiple attack vectors, a detection that relies on 2 or 4 packets is not only accurate, but also achieves dramatic time-saving, e.g., an average time-saving of 99.79\% in terms of flow durations using CICDDoS2019.

\bibliographystyle{IEEEtran}
\bibliography{ddos_settree}

\begin{thebibliography}{10}
\providecommand{\url}[1]{#1}
\csname url@samestyle\endcsname
\providecommand{\newblock}{\relax}
\providecommand{\bibinfo}[2]{#2}
\providecommand{\BIBentrySTDinterwordspacing}{\spaceskip=0pt\relax}
\providecommand{\BIBentryALTinterwordstretchfactor}{4}
\providecommand{\BIBentryALTinterwordspacing}{\spaceskip=\fontdimen2\font plus
\BIBentryALTinterwordstretchfactor\fontdimen3\font minus
  \fontdimen4\font\relax}
\providecommand{\BIBforeignlanguage}[2]{{%
\expandafter\ifx\csname l@#1\endcsname\relax
\typeout{** WARNING: IEEEtran.bst: No hyphenation pattern has been}%
\typeout{** loaded for the language `#1'. Using the pattern for}%
\typeout{** the default language instead.}%
\else
\language=\csname l@#1\endcsname
\fi
#2}}
\providecommand{\BIBdecl}{\relax}
\BIBdecl

\bibitem{07366989709452305}
\BIBentryALTinterwordspacing
B.~Menkus, ``Understanding the denial of service threat,'' \emph{EDPACS},
  vol.~24, no.~9, pp. 11--17, 1997. [Online]. Available:
  \url{https://doi.org/10.1080/07366989709452305}
\BIBentrySTDinterwordspacing

\bibitem{Kaspersky2022}
\BIBentryALTinterwordspacing
Kaspersky, ``Kaspersky: {DDoS} attacks in q4 2021,'' Feb 2022. [Online].
  Available: \url{https://securelist.com/ddos-attacks-in-q4-2021/105784}
\BIBentrySTDinterwordspacing

\bibitem{Radware2022}
\BIBentryALTinterwordspacing
Radware, ``2021 year in review: Denial of service,'' Jan 2022. [Online].
  Available:
  \url{https://blog.radware.com/security/ddos/2022/01/2021-year-in-review-denial-of-service}
\BIBentrySTDinterwordspacing

\bibitem{roman2018mobile}
R.~Roman, J.~Lopez, and M.~Mambo, ``Mobile edge computing, fog et al.: A survey
  and analysis of security threats and challenges,'' \emph{Future Generation
  Computer Systems}, vol.~78, pp. 680--698, 2018.

\bibitem{vignau201910}
B.~Vignau, R.~Khoury, and S.~Hall{\'e}, ``10 years of {IoT} malware: A
  feature-based taxonomy,'' in \emph{IEEE 19th International Conference on
  Software Quality, Reliability and Security Companion (QRS-C)}.\hskip 1em plus
  0.5em minus 0.4em\relax IEEE, 2019, pp. 458--465.

\bibitem{xiao2018security}
L.~Xiao, X.~Wan, C.~Dai, X.~Du, X.~Chen, and M.~Guizani, ``Security in mobile
  edge caching with reinforcement learning,'' \emph{IEEE Wireless
  Communications}, vol.~25, no.~3, pp. 116--122, 2018.

\bibitem{antonakakis2017understanding}
M.~Antonakakis, T.~April, M.~Bailey, M.~Bernhard, E.~Bursztein, J.~Cochran,
  Z.~Durumeric, J.~A. Halderman, L.~Invernizzi, M.~Kallitsis \emph{et~al.},
  ``Understanding the mirai botnet,'' in \emph{26th USENIX security symposium
  (USENIX Security 17)}, 2017, pp. 1093--1110.

\bibitem{arshi2020survey}
M.~Arshi, M.~Nasreen, and K.~Madhavi, ``A survey of {DDOS} attacks using
  machine learning techniques,'' in \emph{E3S Web of Conferences}, vol.
  184.\hskip 1em plus 0.5em minus 0.4em\relax EDP Sciences, 2020, p. 01052.

\bibitem{al2020survey}
M.~A. Al-Garadi, A.~Mohamed, A.~K. Al-Ali, X.~Du, I.~Ali, and M.~Guizani, ``A
  survey of machine and deep learning methods for internet of things ({IoT})
  security,'' \emph{IEEE Communications Surveys \& Tutorials}, vol.~22, no.~3,
  pp. 1646--1685, 2020.

\bibitem{mittal2022deep}
M.~Mittal, K.~Kumar, and S.~Behal, ``Deep learning approaches for detecting
  {DDoS} attacks: a systematic review,'' \emph{Soft Computing}, pp. 1--37,
  2022.

\bibitem{roesch1999snort}
M.~Roesch \emph{et~al.}, ``Snort: Lightweight intrusion detection for
  networks.'' in \emph{Lisa}, vol.~99, no.~1, 1999, pp. 229--238.

\bibitem{ficke2018characterizing}
E.~Ficke, K.~M. Schweitzer, R.~M. Bateman, and S.~Xu, ``Characterizing the
  effectiveness of network-based intrusion detection systems,'' in \emph{IEEE
  Military Communications Conference (MILCOM)}.\hskip 1em plus 0.5em minus
  0.4em\relax IEEE, 2018, pp. 76--81.

\bibitem{umer2017flow}
M.~F. Umer, M.~Sher, and Y.~Bi, ``Flow-based intrusion detection: Techniques
  and challenges,'' \emph{Computers \& Security}, vol.~70, pp. 238--254, 2017.

\bibitem{sperotto2010overview}
A.~Sperotto, G.~Schaffrath, R.~Sadre, C.~Morariu, A.~Pras, and B.~Stiller, ``An
  overview of {IP} flow-based intrusion detection,'' \emph{IEEE communications
  surveys \& tutorials}, vol.~12, no.~3, pp. 343--356, 2010.

\bibitem{sperotto2011flow}
A.~Sperotto and A.~Pras, ``Flow-based intrusion detection,'' in \emph{12th
  IFIP/IEEE International Symposium on Integrated Network Management (IM 2011)
  and Workshops}.\hskip 1em plus 0.5em minus 0.4em\relax IEEE, 2011, pp.
  958--963.

\bibitem{elejla2018flow}
O.~E. Elejla, M.~Anbar, B.~Belaton, and B.~O. Alijla, ``Flow-based {IDS} for
  {ICMPv6-based DDoS} attacks detection,'' \emph{Arabian Journal for Science
  and Engineering}, vol.~43, no.~12, pp. 7757--7775, 2018.

\bibitem{pascoal2020slow}
T.~A. Pascoal, I.~E. Fonseca, and V.~Nigam, ``Slow denial-of-service attacks on
  software defined networks,'' \emph{Computer Networks}, vol. 173, p. 107223,
  2020.

\bibitem{pmlr-v139-hirsch21a}
\BIBentryALTinterwordspacing
R.~Hirsch and R.~Gilad-Bachrach, ``Trees with attention for set prediction
  tasks,'' in \emph{Proceedings of the 38th International Conference on Machine
  Learning}, ser. Proceedings of Machine Learning Research, M.~Meila and
  T.~Zhang, Eds., vol. 139.\hskip 1em plus 0.5em minus 0.4em\relax PMLR, 18--24
  Jul 2021, pp. 4250--4261. [Online]. Available:
  \url{https://proceedings.mlr.press/v139/hirsch21a.html}
\BIBentrySTDinterwordspacing

\bibitem{abeshu2018deep}
A.~Abeshu and N.~Chilamkurti, ``Deep learning: the frontier for distributed
  attack detection in fog-to-things computing,'' \emph{IEEE Communications
  Magazine}, vol.~56, no.~2, pp. 169--175, 2018.

\bibitem{sharafaldin2019developing}
I.~Sharafaldin, A.~H. Lashkari, S.~Hakak, and A.~A. Ghorbani, ``Developing
  realistic distributed denial of service ({DDoS}) attack dataset and
  taxonomy,'' in \emph{2019 International Carnahan Conference on Security
  Technology (ICCST)}.\hskip 1em plus 0.5em minus 0.4em\relax IEEE, 2019, pp.
  1--8.

\bibitem{CIC1_hussain2020network}
\BIBentryALTinterwordspacing
Y.~S. Hussain, ``Network intrusion detection for distributed denial-of-service
  ({DDoS}) attacks using machine learning classification techniques,''
  University of Victoria, Tech. Rep., 2020. [Online]. Available:
  \url{http://hdl.handle.net/1828/11679}
\BIBentrySTDinterwordspacing

\bibitem{CIC2_ortet2021towards}
I.~Ortet~Lopes, D.~Zou, F.~A. Ruambo, S.~Akbar, and B.~Yuan, ``Towards
  effective detection of recent {DDoS} attacks: A deep learning approach,''
  \emph{Security and Communication Networks}, vol. 2021, 2021.

\bibitem{CIC3_elsayed2020ddosnet}
M.~S. Elsayed, N.-A. Le-Khac, S.~Dev, and A.~D. Jurcut, ``Ddosnet: A
  deep-learning model for detecting network attacks,'' in \emph{2020 IEEE 21st
  International Symposium on" A World of Wireless, Mobile and Multimedia
  Networks"(WoWMoM)}.\hskip 1em plus 0.5em minus 0.4em\relax IEEE, 2020, pp.
  391--396.

\bibitem{CIC4_li2020rtvd}
J.~Li, M.~Liu, Z.~Xue, X.~Fan, and X.~He, ``{RTVD}: A real-time volumetric
  detection scheme for {DDoS} in the internet of things,'' \emph{IEEE Access},
  vol.~8, pp. 36\,191--36\,201, 2020.

\bibitem{CIC5_novaes2020long}
M.~P. Novaes, L.~F. Carvalho, J.~Lloret, and M.~L. Proenca, ``Long short-term
  memory and fuzzy logic for anomaly detection and mitigation in
  software-defined network environment,'' \emph{IEEE Access}, vol.~8, pp.
  83\,765--83\,781, 2020.

\bibitem{CIC6_jia2020flowguard}
Y.~Jia, F.~Zhong, A.~Alrawais, B.~Gong, and X.~Cheng, ``Flowguard: an
  intelligent edge defense mechanism against {IoT DDoS} attacks,'' \emph{IEEE
  Internet of Things Journal}, vol.~7, no.~10, pp. 9552--9562, 2020.

\bibitem{CIC7_rajagopal2021towards}
S.~Rajagopal, P.~P. Kundapur, and K.~Hareesha, ``Towards effective network
  intrusion detection: from concept to creation on azure cloud,'' \emph{IEEE
  Access}, vol.~9, pp. 19\,723--19\,742, 2021.

\bibitem{CIC8_almiani2021ddos}
M.~Almiani, A.~AbuGhazleh, Y.~Jararweh, and A.~Razaque, ``{DDoS} detection in
  {5G-enabled IoT} networks using deep {Kalman} backpropagation neural
  network,'' \emph{International Journal of Machine Learning and Cybernetics},
  vol.~12, no.~11, pp. 3337--3349, 2021.

\bibitem{CIC9_de2020near}
M.~V. de~Assis, L.~F. Carvalho, J.~J. Rodrigues, J.~Lloret, and M.~L.
  Proen{\c{c}}a~Jr, ``Near real-time security system applied to sdn
  environments in iot networks using convolutional neural network,''
  \emph{Computers \& Electrical Engineering}, vol.~86, p. 106738, 2020.

\bibitem{roopak2019deep}
M.~Roopak, G.~Y. Tian, and J.~Chambers, ``Deep learning models for cyber
  security in {IoT} networks,'' in \emph{9th annual computing and communication
  workshop and conference (CCWC)}.\hskip 1em plus 0.5em minus 0.4em\relax IEEE,
  2019, pp. 0452--0457.

\bibitem{doriguzzi2020lucid}
R.~Doriguzzi-Corin, S.~Millar, S.~Scott-Hayward, J.~Martinez-del Rincon, and
  D.~Siracusa, ``{LUCID}: A practical, lightweight deep learning solution for
  {DDoS} attack detection,'' \emph{IEEE Transactions on Network and Service
  Management}, vol.~17, no.~2, pp. 876--889, 2020.

\bibitem{belarbi2022intrusion}
O.~Belarbi, A.~Khan, P.~Carnelli, and T.~Spyridopoulos, ``An intrusion
  detection system based on deep belief networks,'' in \emph{Science of Cyber
  Security: 4th International Conference, SciSec 2022, Matsue, Japan, August
  10--12, 2022, Revised Selected Papers}.\hskip 1em plus 0.5em minus
  0.4em\relax Springer, 2022, pp. 377--392.

\bibitem{feinstein2003statistical}
L.~Feinstein, D.~Schnackenberg, R.~Balupari, and D.~Kindred, ``Statistical
  approaches to {DDoS} attack detection and response,'' in \emph{Proceedings
  DARPA information survivability conference and exposition}, vol.~1.\hskip 1em
  plus 0.5em minus 0.4em\relax IEEE, 2003, pp. 303--314.

\bibitem{bojovic2019practical}
P.~Bojovi{\'c}, I.~Ba{\v{s}}i{\v{c}}evi{\'c}, S.~Ocovaj, and M.~Popovi{\'c},
  ``A practical approach to detection of distributed denial-of-service attacks
  using a hybrid detection method,'' \emph{Computers \& Electrical
  Engineering}, vol.~73, pp. 84--96, 2019.

\bibitem{subbulakshmi2011detection}
T.~Subbulakshmi, K.~BalaKrishnan, S.~M. Shalinie, D.~AnandKumar,
  V.~GanapathiSubramanian, and K.~Kannathal, ``Detection of {DDoS} attacks
  using enhanced support vector machines with real time generated dataset,'' in
  \emph{2011 Third International Conference on Advanced Computing}.\hskip 1em
  plus 0.5em minus 0.4em\relax IEEE, 2011, pp. 17--22.

\bibitem{ertoz2004minds}
L.~Ertoz, E.~Eilertson, A.~Lazarevic, P.-N. Tan, V.~Kumar, J.~Srivastava, and
  P.~Dokas, ``Minds-minnesota intrusion detection system,'' \emph{Next
  generation data mining}, pp. 199--218, 2004.

\bibitem{qin2004frequent}
M.~Qin and K.~Hwang, ``Frequent episode rules for intrusive anomaly detection
  with internet datamining,'' in \emph{USENIX Security Symposium}, 2004, pp.
  1--15.

\bibitem{chen2018detection}
L.~Chen, Y.~Zhang, Q.~Zhao, G.~Geng, and Z.~Yan, ``Detection of dns ddos
  attacks with random forest algorithm on spark,'' \emph{Procedia computer
  science}, vol. 134, pp. 310--315, 2018.

\bibitem{chen2018xgboost}
Z.~Chen, F.~Jiang, Y.~Cheng, X.~Gu, W.~Liu, and J.~Peng, ``Xgboost classifier
  for ddos attack detection and analysis in sdn-based cloud,'' in \emph{IEEE
  international conference on big data and smart computing (bigcomp)}.\hskip
  1em plus 0.5em minus 0.4em\relax IEEE, 2018, pp. 251--256.

\bibitem{doshi2018machine}
R.~Doshi, N.~Apthorpe, and N.~Feamster, ``Machine learning {DDoS} detection for
  consumer internet of things devices,'' in \emph{IEEE Security and Privacy
  Workshops (SPW)}.\hskip 1em plus 0.5em minus 0.4em\relax IEEE, 2018, pp.
  29--35.

\bibitem{hping3}
\BIBentryALTinterwordspacing
S.~Sanfilippo, ``hping3(8) - linux man page,'' Feb 2022. [Online]. Available:
  \url{https://linux.die.net/man/8/hping3}
\BIBentrySTDinterwordspacing

\bibitem{sharafaldin2018toward}
I.~Sharafaldin, A.~H. Lashkari, and A.~A. Ghorbani, ``Toward generating a new
  intrusion detection dataset and intrusion traffic characterization.''
  \emph{ICISSp}, vol.~1, pp. 108--116, 2018.

\bibitem{caida}
\BIBentryALTinterwordspacing
CAIDA, ``The {CAIDA UCSD} ``{DDoS Attack} 2007'' dataset,'' 2007. [Online].
  Available: \url{http://www.caida.org/data/passive/ddos-20070804 dataset.xml.}
\BIBentrySTDinterwordspacing

\bibitem{darpa}
\BIBentryALTinterwordspacing
DARPA, ``{DARPA} 2000 intrustion detection scenario specific data sets,'' 2000.
  [Online]. Available: \url{https://www.ll.mit.edu/r-d/datasets/
  2000-darpa-intrusion-detection-scenario-specific-data-sets}
\BIBentrySTDinterwordspacing

\bibitem{brown2009analysis}
C.~Brown, A.~Cowperthwaite, A.~Hijazi, and A.~Somayaji, ``Analysis of the 1999
  {DARPA/Lincoln} laboratory {IDS} evaluation data with netadhict,'' in
  \emph{2009 IEEE Symposium on Computational Intelligence for Security and
  Defense Applications}.\hskip 1em plus 0.5em minus 0.4em\relax IEEE, 2009, pp.
  1--7.

\bibitem{singh2015approach}
K.~J. Singh and T.~De, ``An approach of {DDoS} attack detection using
  classifiers,'' in \emph{Emerging Research in Computing, Information,
  Communication and Applications}.\hskip 1em plus 0.5em minus 0.4em\relax
  Springer, 2015, pp. 429--437.

\bibitem{yu2011discriminating}
S.~Yu, W.~Zhou, W.~Jia, S.~Guo, Y.~Xiang, and F.~Tang, ``Discriminating {DDoS}
  attacks from flash crowds using flow correlation coefficient,'' \emph{IEEE
  transactions on parallel and distributed systems}, vol.~23, no.~6, pp.
  1073--1080, 2011.

\bibitem{verkerken2023novel}
M.~Verkerken, L.~D’hooge, D.~Sudyana, Y.-D. Lin, T.~Wauters, B.~Volckaert,
  and F.~De~Turck, ``A novel multi-stage approach for hierarchical intrusion
  detection,'' \emph{IEEE Transactions on Network and Service Management},
  2023.

\bibitem{bernaille2006traffic}
L.~Bernaille, R.~Teixeira, I.~Akodkenou, A.~Soule, and K.~Salamatian, ``Traffic
  classification on the fly,'' \emph{ACM SIGCOMM Computer Communication
  Review}, vol.~36, no.~2, pp. 23--26, 2006.

\bibitem{li2008real}
J.~Li, S.~Zhang, Y.~Lu, and J.~Yan, ``Real-time p2p traffic identification,''
  in \emph{IEEE GLOBECOM 2008-2008 IEEE Global Telecommunications
  Conference}.\hskip 1em plus 0.5em minus 0.4em\relax IEEE, 2008, pp. 1--5.

\bibitem{gu2011realtime}
C.~Gu, S.~Zhang, and Y.~Sun, ``Realtime encrypted traffic identification using
  machine learning.'' \emph{J. Softw.}, vol.~6, no.~6, pp. 1009--1016, 2011.

\bibitem{liu2017novel}
Y.~Liu, J.~Chen, P.~Chang, and X.~Yun, ``A novel algorithm for encrypted
  traffic classification based on sliding window of flow's first n packets,''
  in \emph{2017 2nd IEEE International Conference on Computational Intelligence
  and Applications (ICCIA)}.\hskip 1em plus 0.5em minus 0.4em\relax IEEE, 2017,
  pp. 463--470.

\bibitem{NEURIPS2022_0378c769}
\BIBentryALTinterwordspacing
L.~Grinsztajn, E.~Oyallon, and G.~Varoquaux, ``Why do tree-based models still
  outperform deep learning on typical tabular data?'' in \emph{Advances in
  Neural Information Processing Systems}, S.~Koyejo, S.~Mohamed, A.~Agarwal,
  D.~Belgrave, K.~Cho, and A.~Oh, Eds., vol.~35.\hskip 1em plus 0.5em minus
  0.4em\relax Curran Associates, Inc., 2022, pp. 507--520. [Online]. Available:
  \url{https://proceedings.neurips.cc/paper_files/paper/2022/file/0378c7692da36807bdec87ab043cdadc-Paper-Datasets_and_Benchmarks.pdf}
\BIBentrySTDinterwordspacing

\bibitem{NetFlow}
\BIBentryALTinterwordspacing
Cisco, ``Netflow version 9 flow-record format,'' Jan 2011. [Online]. Available:
  \url{https://www.cisco.com/en/US/technologies/tk648/tk362/technologies_white_paper09186a00800a3db9.html}
\BIBentrySTDinterwordspacing

\bibitem{quiring2022and}
E.~Quiring, F.~Pendlebury, A.~Warnecke, F.~Pierazzi, C.~Wressnegger,
  L.~Cavallaro, and K.~Rieck, ``Dos and don’ts of machine learning in
  computer security,'' in \emph{31st USENIX Security Symposium (USENIX Security
  22), USENIX Association, Boston, MA}, 2022.

\bibitem{ahmim2019novel}
A.~Ahmim, L.~Maglaras, M.~A. Ferrag, M.~Derdour, and H.~Janicke, ``A novel
  hierarchical intrusion detection system based on decision tree and
  rules-based models,'' in \emph{2019 15th International Conference on
  Distributed Computing in Sensor Systems (DCOSS)}.\hskip 1em plus 0.5em minus
  0.4em\relax Santorini Island, Greece: IEEE, 2019, pp. 228--233.

\end{thebibliography}

\newpage
\appendices

\section{Flows duration distribution per attack vector}
\label{appendix:duration_attack_vectors}

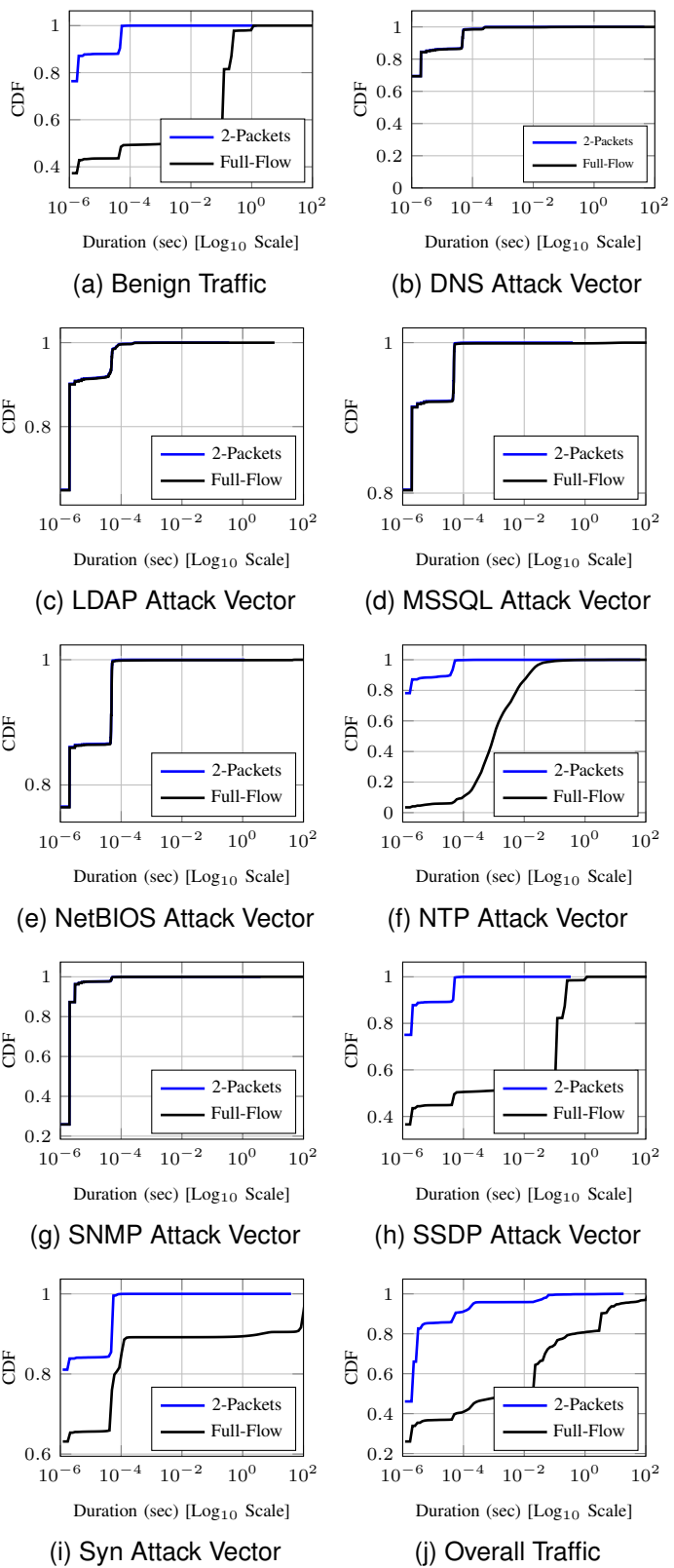
\begin{figure}
   \centering
     \subfloat[Benign Traffic]{
         \begin{tikzpicture}
         \begin{axis}[    xlabel={Duration (sec) [Log$_{10}$ Scale]},
             height=4cm,
             width=4.9cm,
             ylabel={CDF},
             ylabel style={yshift=-15pt},
             ylabel style={font=\scriptsize},
             yticklabel style={font=\scriptsize}, 
             xlabel style={font=\scriptsize}, 
             xticklabel style={font=\scriptsize},
             xtick={-6,-4,-2, 0, 2},
             ytick distance=0.2,
             xticklabels={$10^{-6}$, $10^{-4}$, $10^{-2}$, $10^{0}$, $10^{2}$},
             xmin=-6,
             xmax=2,
             grid=both,
             legend style={font=\scriptsize},
             legend pos=south east,
             ]
         \addplot[blue, line width=1pt] table[x index=0,y index=1,col sep=comma] {csv/ddos/cdf/DrDoS_UDP_2packet_flow_log.csv};  
         \addplot[line width=1pt] table[x index=0,y index=1,col sep=comma] {csv/ddos/cdf/DrDoS_UDP_full_flow_log.csv}; 
         \legend{2-Packets, Full-Flow}
         \end{axis}
         \end{tikzpicture} 
    }
    \subfloat[DNS Attack Vector]{%
         \begin{tikzpicture}
         \begin{axis}[xlabel={Duration (sec) [Log$_{10}$ Scale]},
             height=4cm,
             width=4.9cm,
             ylabel={CDF},
             ylabel style={yshift=-15pt},
             ylabel style={font=\scriptsize},
             yticklabel style={font=\scriptsize}, 
             xlabel style={font=\scriptsize}, 
             xticklabel style={font=\scriptsize}, 
             xtick={-6,-4,-2, 0, 2},
             xticklabels={$10^{-6}$, $10^{-4}$, $10^{-2}$, $10^{0}$, $10^{2}$},
             xmin=-6,
             xmax=2,
             ymin=0,
             grid=both,
             ytick distance=0.2,
             legend style={font=\tiny},
             legend pos=south east,
             ]
         \addplot[line width=1pt, blue] table[x index=0,y index=1,col sep=comma] {csv/ddos/cdf/DrDoS_DNS_2packet_flow_log.csv};  
         \addplot[line width=1pt] table[x index=0,y index=1,col sep=comma] {csv/ddos/cdf/DrDoS_DNS_full_flow_log.csv}; 
         \legend{2-Packets, Full-Flow}
         \end{axis}
         \end{tikzpicture} 
         }\\
    \subfloat[LDAP Attack Vector]{%
        \begin{tikzpicture}
        \begin{axis}[xlabel={Duration (sec) [Log$_{10}$ Scale]},
            height=4cm,
            width=4.9cm,
            ylabel={CDF},
            ylabel style={yshift=-15pt},
            ylabel style={font=\scriptsize},
            yticklabel style={font=\scriptsize}, 
            xlabel style={font=\scriptsize}, 
            xticklabel style={font=\scriptsize},
            xtick={-6,-4,-2, 0, 2},
            xticklabels={$10^{-6}$, $10^{-4}$, $10^{-2}$, $10^{0}$, $10^{2}$},
            xmin=-6,
            xmax=2,
            ytick distance=0.2,
            grid=both,
            legend style={font=\scriptsize},
            legend pos=south east,
            ]
        \addplot[line width=1pt, blue] table[x index=0,y index=1,col sep=comma] {csv/ddos/cdf/DrDoS_LDAP_2packet_flow_log.csv};  
        \addplot[line width=1pt] table[x index=0,y index=1,col sep=comma] {csv/ddos/cdf/DrDoS_LDAP_full_flow_log.csv}; 
        \legend{2-Packets, Full-Flow}
        \end{axis}
        \end{tikzpicture} 
    }
    \subfloat[MSSQL Attack Vector]{%
        \begin{tikzpicture}
        \begin{axis}[    xlabel={Duration (sec) [Log$_{10}$ Scale]},
            height=4cm,
            width=4.9cm,
            ylabel={CDF},
            ylabel style={yshift=-15pt},
            ylabel style={font=\scriptsize},
            yticklabel style={font=\scriptsize}, 
            xlabel style={font=\scriptsize}, 
            xticklabel style={font=\scriptsize}, 
            xtick={-6,-4,-2, 0, 2},
            xticklabels={$10^{-6}$, $10^{-4}$, $10^{-2}$, $10^{0}$, $10^{2}$},
            xmin=-6,
            ytick distance=0.2,
            xmax=2,
            grid=both,
            legend style={font=\scriptsize},
            legend pos=south east,
            ]
        \addplot[line width=1pt, blue] table[x index=0,y index=1,col sep=comma] {csv/ddos/cdf/DrDoS_MSSQL_2packet_flow_log.csv};  
        \addplot[line width=1pt] table[x index=0,y index=1,col sep=comma] {csv/ddos/cdf/DrDoS_MSSQL_full_flow_log.csv}; 
        \legend{2-Packets, Full-Flow}
        \end{axis}
        \end{tikzpicture} 
}\\
    \subfloat[NetBIOS Attack Vector]{%
        \begin{tikzpicture}
        \begin{axis}[    xlabel={Duration (sec) [Log$_{10}$ Scale]},
            height=4cm,
            width=4.9cm,
            ylabel={CDF},
            ylabel style={yshift=-15pt},
            ylabel style={font=\scriptsize},
            yticklabel style={font=\scriptsize}, 
            xlabel style={font=\scriptsize}, 
            xticklabel style={font=\scriptsize},
            xtick={-6,-4,-2, 0, 2},
            ytick distance=0.2,
            xticklabels={$10^{-6}$, $10^{-4}$, $10^{-2}$, $10^{0}$, $10^{2}$},
            xmin=-6,
            xmax=2,
            grid=both,
            legend style={font=\scriptsize},
            legend pos=south east,
            ]
        \addplot[blue, line width=1pt] table[x index=0,y index=1,col sep=comma] {csv/ddos/cdf/DrDoS_NetBIOS_2packet_flow_log.csv};  
        \addplot[line width=1pt] table[x index=0,y index=1,col sep=comma] {csv/ddos/cdf/DrDoS_NetBIOS_full_flow_log.csv}; 
        \legend{2-Packets, Full-Flow}
        \end{axis}
        \end{tikzpicture} 
}
\subfloat[NTP Attack Vector]{%
        \begin{tikzpicture}
        \begin{axis}[    xlabel={Duration (sec) [Log$_{10}$ Scale]},
            height=4cm,
            width=4.9cm,
            ylabel={CDF},
            ylabel style={yshift=-15pt},
            ylabel style={font=\scriptsize},
            yticklabel style={font=\scriptsize}, 
            xlabel style={font=\scriptsize}, 
            xticklabel style={font=\scriptsize},
            xtick={-6,-4,-2, 0, 2},
            ytick distance=0.2,
            xticklabels={$10^{-6}$, $10^{-4}$, $10^{-2}$, $10^{0}$, $10^{2}$},
            xmin=-6,
            xmax=2,
            grid=both,
            legend style={font=\scriptsize},
            legend pos=south east,
            ]
        \addplot[blue, line width=1pt] table[x index=0,y index=1,col sep=comma] {csv/ddos/cdf/DrDoS_NTP_2packet_flow_log.csv};  
        \addplot[line width=1pt] table[x index=0,y index=1,col sep=comma] {csv/ddos/cdf/DrDoS_NTP_full_flow_log.csv}; 
        \legend{2-Packets, Full-Flow}
        \end{axis}
        \end{tikzpicture} 
}\\
\subfloat[SNMP Attack Vector]{%
        \begin{tikzpicture}
        \begin{axis}[    xlabel={Duration (sec) [Log$_{10}$ Scale]},
            height=4cm,
            width=4.9cm,
            ylabel={CDF},
            ylabel style={yshift=-15pt},
            ylabel style={font=\scriptsize},
            yticklabel style={font=\scriptsize}, 
            xlabel style={font=\scriptsize}, 
            xticklabel style={font=\scriptsize},
            xtick={-6,-4,-2, 0, 2},
            ytick distance=0.2,
            xticklabels={$10^{-6}$, $10^{-4}$, $10^{-2}$, $10^{0}$, $10^{2}$},
            xmin=-6,
            xmax=2,
            grid=both,
            legend style={font=\scriptsize},
            legend pos=south east,
            ]
        \addplot[blue, line width=1pt] table[x index=0,y index=1,col sep=comma] {csv/ddos/cdf/DrDoS_SNMP_2packet_flow_log.csv};  
        \addplot[line width=1pt] table[x index=0,y index=1,col sep=comma] {csv/ddos/cdf/DrDoS_SNMP_full_flow_log.csv}; 
        \legend{2-Packets, Full-Flow}
        \end{axis}
        \end{tikzpicture} 
}
\subfloat[SSDP Attack Vector]{%
        \begin{tikzpicture}
        \begin{axis}[    xlabel={Duration (sec) [Log$_{10}$ Scale]},
            height=4cm,
            width=4.9cm,
            ylabel={CDF},
            ylabel style={yshift=-15pt},
            ylabel style={font=\scriptsize},
            yticklabel style={font=\scriptsize}, 
            xlabel style={font=\scriptsize}, 
            xticklabel style={font=\scriptsize},
            xtick={-6,-4,-2, 0, 2},
            ytick distance=0.2,
            xticklabels={$10^{-6}$, $10^{-4}$, $10^{-2}$, $10^{0}$, $10^{2}$},
            xmin=-6,
            xmax=2,
            grid=both,
            legend style={font=\scriptsize},
            legend pos=south east,
            ]
        \addplot[blue, line width=1pt] table[x index=0,y index=1,col sep=comma] {csv/ddos/cdf/DrDoS_SSDP_2packet_flow_log.csv};  
        \addplot[line width=1pt] table[x index=0,y index=1,col sep=comma] {csv/ddos/cdf/DrDoS_SSDP_full_flow_log.csv}; 
        \legend{2-Packets, Full-Flow}
        \end{axis}
        \end{tikzpicture} 
}\\
\subfloat[Syn Attack Vector]{%
        \begin{tikzpicture}
        \begin{axis}[    xlabel={Duration (sec) [Log$_{10}$ Scale]},
            height=4cm,
            width=4.9cm,
            ylabel={CDF},
            ylabel style={yshift=-15pt},
            ylabel style={font=\scriptsize},
            yticklabel style={font=\scriptsize}, 
            xlabel style={font=\scriptsize}, 
            xticklabel style={font=\scriptsize},
            xtick={-6,-4,-2, 0, 2},
            ytick distance=0.2,
            xticklabels={$10^{-6}$, $10^{-4}$, $10^{-2}$, $10^{0}$, $10^{2}$},
            xmin=-6,
            xmax=2,
            grid=both,
            legend style={font=\scriptsize},
            legend pos=south east,
            ]
        \addplot[blue, line width=1pt] table[x index=0,y index=1,col sep=comma] {csv/ddos/cdf/Syn_2packet_flow_log.csv};  
        \addplot[line width=1pt] table[x index=0,y index=1,col sep=comma] {csv/ddos/cdf/Syn_full_flow_log.csv}; 
        \legend{2-Packets, Full-Flow}
        \end{axis}
        \end{tikzpicture} 
}
\subfloat[Overall Traffic]{%
        \begin{tikzpicture}
        \begin{axis}[    xlabel={Duration (sec) [Log$_{10}$ Scale]},
            height=4cm,
            width=4.9cm,
            ylabel={CDF},
            ylabel style={yshift=-15pt},
            ylabel style={font=\scriptsize},
            yticklabel style={font=\scriptsize}, 
            xlabel style={font=\scriptsize}, 
            xticklabel style={font=\scriptsize},
            xtick={-6,-4,-2, 0, 2},
            ytick distance=0.2,
            xticklabels={$10^{-6}$, $10^{-4}$, $10^{-2}$, $10^{0}$, $10^{2}$},
            xmin=-6,
            xmax=2,
            grid=both,
            legend style={font=\scriptsize},
            legend pos=south east,
            ]
        \addplot[blue, line width=1pt] table[x index=0,y index=1,col sep=comma] {csv/ddos/cdf/overall_2packet_flow_log.csv};  
        \addplot[line width=1pt] table[x index=0,y index=1,col sep=comma] {csv/ddos/cdf/overall_full_flow_log.csv}; 
        \legend{2-Packets, Full-Flow}
        \end{axis}
        \end{tikzpicture} 
}  
   \caption{CICDDoS2019 flow duration distribution - a comparison between first 2-packets and complete flows.}
   \label{fig:ddos2019_duration_all_vectors}
\end{figure}

\begin{figure}
\vspace{0.7cm}
\subfloat[Bot Attack Vector]{%
        \begin{tikzpicture}
        \begin{axis}[xlabel={Duration (sec) [Log$_{10}$ Scale]},
            height=4cm,
            width=4.9cm,
            ylabel={CDF},
            ylabel style={yshift=-15pt},
            ylabel style={font=\scriptsize},
            yticklabel style={font=\scriptsize}, 
            ytick distance=0.2,
            xlabel style={font=\scriptsize}, 
            xticklabel style={font=\scriptsize}, 
            xtick={-6,-4,-2, 0, 2},
            xticklabels={$10^{-6}$, $10^{-4}$, $10^{-2}$, $10^{0}$, $10^{2}$},
            xmin=-6,
            xmax=2,
            grid=both,
            ]
        \addplot[color10, line width=1pt] table[x index=0,y index=1,col sep=comma] {csv/ids/cdf/Bot_2packet_flow_log.csv}; 
        \addplot[blue, line width=1pt] table[x index=0,y index=1,col sep=comma] {csv/ids/cdf/Bot_4packet_flow_log.csv}; 
        \addplot[color2, line width=1pt] table[x index=0,y index=1,col sep=comma] {csv/ids/cdf/Bot_full_flow_log.csv}; 
        \end{axis}
        \end{tikzpicture} 
}
\subfloat[DDoS Attack Vector]{%
        \begin{tikzpicture}
        \begin{axis}[xlabel={Duration (sec) [Log$_{10}$ Scale]},
            height=4cm,
            width=4.9cm,
            ylabel={CDF},
            ylabel style={yshift=-15pt},
            ylabel style={font=\scriptsize},
            yticklabel style={font=\scriptsize}, 
            xlabel style={font=\scriptsize}, 
            xticklabel style={font=\scriptsize},
            xtick={-6,-4,-2, 0, 2},
            xticklabels={$10^{-6}$, $10^{-4}$, $10^{-2}$, $10^{0}$, $10^{2}$},
            xmin=-6,
            xmax=2,
            ytick distance=0.2,
            grid=both,
                legend style={at={(0.5,-0.2)},    
                    anchor=north,legend columns=4},  
            ]
        \addplot[color10, line width=1pt] table[x index=0,y index=1,col sep=comma] {csv/ids/cdf/DDoS_2packet_flow_log.csv};  
        \addplot[blue, line width=1pt] table[x index=0,y index=1,col sep=comma] {csv/ids/cdf/DDoS_4packet_flow_log.csv}; 
        \addplot[color2, line width=1pt] table[x index=0,y index=1,col sep=comma] {csv/ids/cdf/DDoS_full_flow_log.csv}; 
        \end{axis}
        \end{tikzpicture} 
}\\
\subfloat[SlowHTTPTest Vector]{%
        \begin{tikzpicture}
        \begin{axis}[    xlabel={Duration (sec) [Log$_{10}$ Scale]},
            height=4cm,
            width=4.9cm,
            ylabel={CDF},
            ylabel style={yshift=-15pt},
            ylabel style={font=\scriptsize},
            yticklabel style={font=\scriptsize}, 
            xlabel style={font=\scriptsize}, 
            xticklabel style={font=\scriptsize}, 
            xtick={-6,-4,-2, 0, 2},
            xticklabels={$10^{-6}$, $10^{-4}$, $10^{-2}$, $10^{0}$, $10^{2}$},
            xmin=-6,
            xmax=2,
            ytick distance=0.2,
            grid=both,
                legend style={at={(0.5,-0.2)},    
                    anchor=north,legend columns=4},  
            ]
        \addplot[color10, line width=1pt] table[x index=0,y index=1,col sep=comma] {csv/ids/cdf/DoS_Slowhttptest_2packet_flow_log.csv};  
        \addplot[blue, line width=1pt] table[x index=0,y index=1,col sep=comma] {csv/ids/cdf/DoS_Slowhttptest_4packet_flow_log.csv}; 
        \addplot[color2, line width=1pt] table[x index=0,y index=1,col sep=comma] {csv/ids/cdf/DoS_Slowhttptest_full_flow_log.csv}; 
        \end{axis}
        \end{tikzpicture} 
}
\subfloat[DoS\_Slowloris Vector]{%
        \begin{tikzpicture}
        \begin{axis}[    xlabel={Duration (sec) [Log$_{10}$ Scale]},
            height=4cm,
            width=4.9cm,
            ylabel={CDF},
            ylabel style={yshift=-15pt},
            ylabel style={font=\scriptsize},
            yticklabel style={font=\scriptsize}, 
            xlabel style={font=\scriptsize}, 
            xticklabel style={font=\scriptsize},
            xtick={-6,-4,-2, 0, 2},
            xticklabels={$10^{-6}$, $10^{-4}$, $10^{-2}$, $10^{0}$, $10^{2}$},
            xmin=-6,
            xmax=2,
            ytick distance=0.2,
            grid=both,
                legend style={at={(0.5,-0.2)},    
                    anchor=north,legend columns=4},  
            ]
        \addplot[color10, line width=1pt] table[x index=0,y index=1,col sep=comma] {csv/ids/cdf/DoS_Slowloris_2packet_flow_log.csv};  
        \addplot[blue, line width=1pt] table[x index=0,y index=1,col sep=comma] {csv/ids/cdf/DoS_Slowloris_4packet_flow_log.csv}; 
        \addplot[color2, line width=1pt] table[x index=0,y index=1,col sep=comma] {csv/ids/cdf/DoS_Slowloris_full_flow_log.csv}; 
        \end{axis}
        \end{tikzpicture} 
}\\
\subfloat[FTP-Patator Vector]{%
        \begin{tikzpicture}
        \begin{axis}[    xlabel={Duration (sec) [Log$_{10}$ Scale]},
            height=4cm,
            width=4.9cm,
            ylabel={CDF},
            ylabel style={yshift=-15pt},
            ylabel style={font=\scriptsize},
            yticklabel style={font=\scriptsize}, 
            xlabel style={font=\scriptsize}, 
            xticklabel style={font=\scriptsize}, 
            xtick={-6,-4,-2, 0, 2},
            xticklabels={$10^{-6}$, $10^{-4}$, $10^{-2}$, $10^{0}$, $10^{2}$},
            xmin=-6,
            xmax=2,
            ytick distance=0.2,
            grid=both,
                legend style={at={(0.5,-0.2)},    
                    anchor=north,legend columns=4},  
            ]
        \addplot[color10, line width=1pt] table[x index=0,y index=1,col sep=comma] {csv/ids/cdf/FTP-Patator_2packet_flow_log.csv};  
        \addplot[blue, line width=1pt] table[x index=0,y index=1,col sep=comma] {csv/ids/cdf/FTP-Patator_4packet_flow_log.csv}; 
        \addplot[color2, line width=1pt] table[x index=0,y index=1,col sep=comma] {csv/ids/cdf/FTP-Patator_full_flow_log.csv}; 
        \end{axis}
        \end{tikzpicture} 
}
\subfloat[PortScan Vector]{%
        \begin{tikzpicture}
        \begin{axis}[    xlabel={Duration (sec) [Log$_{10}$ Scale]},
            height=4cm,
            width=4.9cm,
            ylabel={CDF},
            ylabel style={yshift=-15pt},
            ylabel style={font=\scriptsize},
            yticklabel style={font=\scriptsize}, 
            xlabel style={font=\scriptsize}, 
            xticklabel style={font=\scriptsize},
            xtick={-6,-4,-2, 0, 2},
            xticklabels={$10^{-6}$, $10^{-4}$, $10^{-2}$, $10^{0}$, $10^{2}$},
            xmin=-6,
            xmax=2,
            ytick distance=0.2,
            grid=both,
                legend style={at={(0.5,-0.2)},    
                    anchor=north,legend columns=4},  
            ]
        \addplot[color10, line width=1pt] table[x index=0,y index=1,col sep=comma] {csv/ids/cdf/PortScan_2packet_flow_log.csv};  
        \addplot[blue, line width=1pt] table[x index=0,y index=1,col sep=comma] {csv/ids/cdf/PortScan_4packet_flow_log.csv}; 
        \addplot[color2, line width=1pt] table[x index=0,y index=1,col sep=comma] {csv/ids/cdf/PortScan_full_flow_log.csv}; 
        \end{axis}
    \end{tikzpicture} 
}\\
\subfloat[SSH-Patator Vector]{%
        \begin{tikzpicture}
        \begin{axis}[    xlabel={Duration (sec) [Log$_{10}$ Scale]},
            height=4cm,
            width=4.9cm,
            ylabel={CDF},
            ylabel style={yshift=-15pt},
            ylabel style={font=\scriptsize},
            yticklabel style={font=\scriptsize}, 
            xlabel style={font=\scriptsize}, 
            xticklabel style={font=\scriptsize}, 
            xtick={-6,-4,-2, 0, 2},
            xticklabels={$10^{-6}$, $10^{-4}$, $10^{-2}$, $10^{0}$, $10^{2}$},
            xmin=-6,
            xmax=2,
            ytick distance=0.2,
            grid=both,
                legend style={at={(0.5,-0.2)},    
                    anchor=north,legend columns=4},  
            ]
        \addplot[color10, line width=1pt] table[x index=0,y index=1,col sep=comma] {csv/ids/cdf/SSH-Patator_2packet_flow_log.csv};  
        \addplot[blue, line width=1pt] table[x index=0,y index=1,col sep=comma] {csv/ids/cdf/SSH-Patator_4packet_flow_log.csv}; 
        \addplot[color2, line width=1pt] table[x index=0,y index=1,col sep=comma] {csv/ids/cdf/SSH-Patator_full_flow_log.csv}; 
        \end{axis}
        \end{tikzpicture} 
}
\subfloat[Web Attack Brute-Force]{%
        \begin{tikzpicture}
        \begin{axis}[    xlabel={Duration (sec) [Log$_{10}$ Scale]},
            height=4cm,
            width=4.9cm,
            ylabel={CDF},
            ylabel style={yshift=-15pt},
            ylabel style={font=\scriptsize},
            yticklabel style={font=\scriptsize}, 
            xlabel style={font=\scriptsize}, 
            xticklabel style={font=\scriptsize},
            xtick={-6,-4,-2, 0, 2},
            xticklabels={$10^{-6}$, $10^{-4}$, $10^{-2}$, $10^{0}$, $10^{2}$},
            xmin=-6,
            xmax=2,
            ytick distance=0.2,
            grid=both,
                legend style={at={(0.5,-0.2)},    
                    anchor=north,legend columns=4},  
            ]
        \addplot[color10, line width=1pt] table[x index=0,y index=1,col sep=comma] {csv/ids/cdf/WABruteForce_2packet_flow_log.csv};  
        \addplot[blue, line width=1pt] table[x index=0,y index=1,col sep=comma] {csv/ids/cdf/WABruteForce_4packet_flow_log.csv}; 
        \addplot[color2, line width=1pt] table[x index=0,y index=1,col sep=comma] {csv/ids/cdf/WABruteForce_full_flow_log.csv}; 
        \end{axis}
        \end{tikzpicture} 
}\\
\subfloat[Web Attack XSS]{%
         \begin{tikzpicture}
        \begin{axis}[    xlabel={Duration (sec) [Log$_{10}$ Scale]},
            height=4cm,
            width=4.9cm,
            ylabel={CDF},
            ylabel style={yshift=-15pt},
            ylabel style={font=\scriptsize},
            yticklabel style={font=\scriptsize}, 
            xlabel style={font=\scriptsize}, 
            xticklabel style={font=\scriptsize}, 
            xtick={-6,-4,-2, 0, 2},
            xticklabels={$10^{-6}$, $10^{-4}$, $10^{-2}$, $10^{0}$, $10^{2}$},
            xmin=-6,
            xmax=2,
            ytick distance=0.2,
            grid=both,
                legend style={at={(0.5,-0.2)},    
                    anchor=north,legend columns=4},  
            ]
        \addplot[color10, line width=1pt] table[x index=0,y index=1,col sep=comma] {csv/ids/cdf/WAXSS_2packet_flow_log.csv};  
        \addplot[blue, line width=1pt] table[x index=0,y index=1,col sep=comma] {csv/ids/cdf/WAXSS_4packet_flow_log.csv}; 
        \addplot[color2, line width=1pt] table[x index=0,y index=1,col sep=comma] {csv/ids/cdf/WAXSS_full_flow_log.csv}; 
        \end{axis}
        \end{tikzpicture} 
}
\subfloat[Overall Traffic]{%
    \centering
        \begin{tikzpicture}
        \begin{axis}[    xlabel={Duration (sec) [Log$_{10}$ Scale]},
            height=4cm,
            width=5.2cm,
            ylabel={CDF},
            ylabel style={yshift=-15pt},
            ylabel style={font=\scriptsize},
            yticklabel style={font=\scriptsize}, 
            xlabel style={font=\scriptsize}, 
            xticklabel style={font=\scriptsize},
            xtick={-6,-4,-2, 0, 2},
            xticklabels={$10^{-6}$, $10^{-4}$, $10^{-2}$, $10^{0}$, $10^{2}$},
            xmin=-6,
            xmax=2,
            ytick distance=0.2,
            grid=both,
                legend style={at={(0.5,-0.2)},    
                    anchor=north,legend columns=4},  
            ]
        \addplot[color10, line width=1pt] table[x index=0,y index=1,col sep=comma] {csv/ids/cdf/overall_2packet_flow_log.csv};  
        \addplot[blue, line width=1pt] table[x index=0,y index=1,col sep=comma] {csv/ids/cdf/overall_4packet_flow_log.csv}; 
        \addplot[color2, line width=1pt] table[x index=0,y index=1,col sep=comma] {csv/ids/cdf/overall_full_flow_log.csv}; 
        \end{axis}
        \end{tikzpicture} 
}\\
\centering
\begin{tikzpicture}
  \matrix[name=mylegend,anchor=south, draw=black, thin, row sep=0.5em] at (0,0) {
    \node [label=right:{\footnotesize First 2-Packets}] {\protect\raisebox{-0.5ex}{\color{green}\rule{1em}{1pt}}}; & 
    \node [label=right:{\footnotesize First 4-Packets}] {\protect\raisebox{-0.5ex}{\color{blue}\rule{1em}{1pt}}}; & 
    \node [label=right:{\footnotesize Complete-Flow}] {\protect\raisebox{-0.5ex}{\color{red}\rule{1em}{1pt}}}; \\
  };
\end{tikzpicture}
   \caption{CICIDS2017 flow duration distribution - a comparison between initial-packet streams and complete flows.}
   \label{fig:ids2017_duration_all_vectors}
\end{figure}
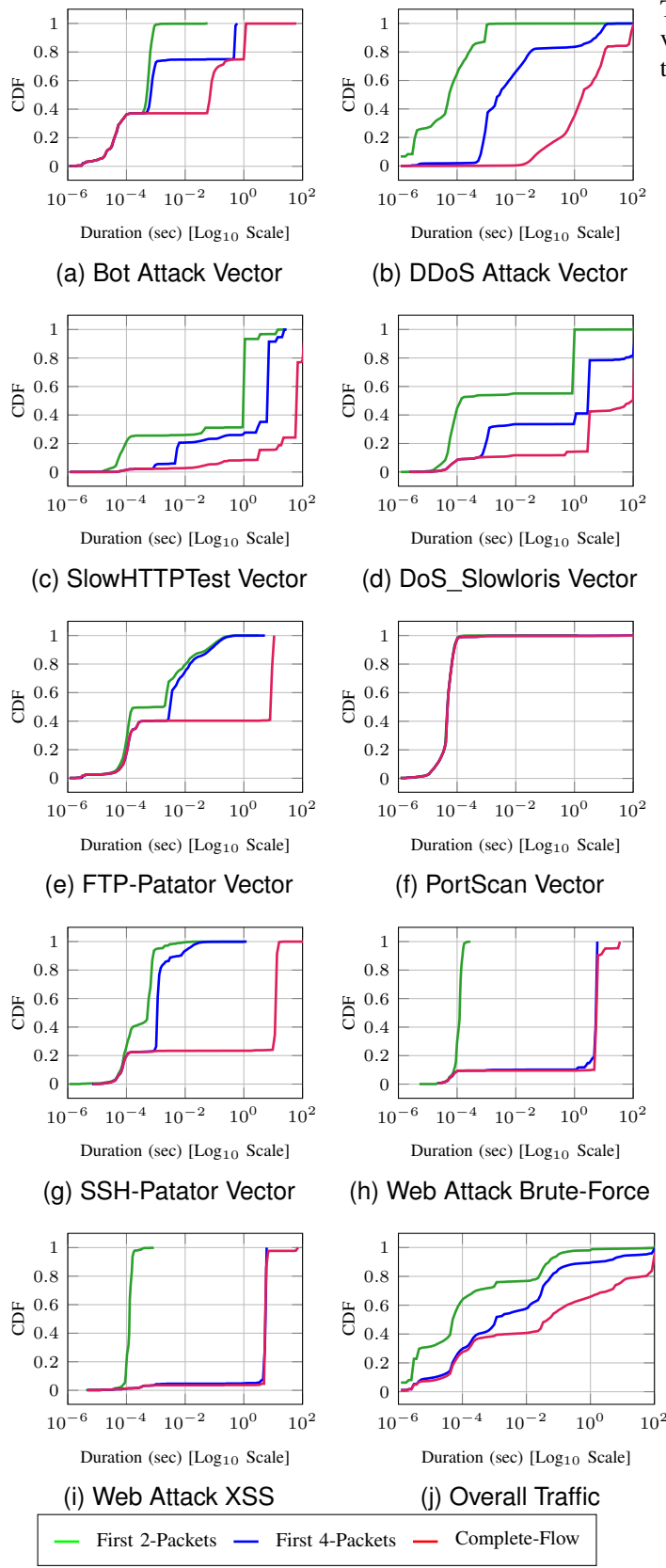

We present the cumulative distribution of flows' duration in both datasets across all the attack vectors.
The distributions of the overall benign and attack traffic were presented in Section \ref{subsec:dataset}, for each dataset, together with example distributions of a few attack vectors. Here, for completeness, we provide the corresponding CDFs for all the attack vectors.
Figures \ref{fig:ddos2019_duration_all_vectors}, \ref{fig:ids2017_duration_all_vectors} present the cumulative distributions of flow durations in CICDDoS2019 and CICIDS2017 attack vectors respectively. 

\section{Evaluation Metrics}
\label{appendix:metrics}
As commonly used to measure Network Intrusion Detection Systems (NIDSs) performance in the literature, we report metrics such as precision, recall, F-score and accuracy to evaluate our proposed model. TP are the true-positives, i.e., attack flows (of any attack vector) that were classified by the model as attacks. FP are the false-positives, i.e., benign flows that were incorrectly classified as attack. Similarly, TN (true negatives) are the benign flows classified as benign, and FN (false negatives) are attack flows incorrectly classified as benign flows.

\textit{Recall} is the ratio of TP over the sum of TP and FN.  It measures the effectiveness of the model in the classification of true attack traffic.
\[Recall=\frac{TP} {TP + FN}.\]     
\textit{Precision} is the ratio of TP over the sum of TP and FP. It measures the percentage of flows classified as attacks, which are true attacks.
\[Precision=\frac{TP} {TP + FP}.\]     
\textit{Accuracy} is the ratio of the sum of TP and FP over the sum of TP, FP, TN and FN. It measures the ratio of correctly classified attack and benign flows over total flows.
\[Accuracy=\frac{TP + TN} {TP + TN + FP + FN}.\]     
\textit{F-score} is the harmonic mean of the \textit{Recall} and the \textit{Precision}.
\[F-score=\]
{\small \[= \frac{2} {{Recall}^{-1}+{Precision}^{-1}} = \frac{2 \times Recall \times Precision} {Recall+Precision}.\]}

\section{Detailed Features Importance}
\label{appendix:features}
\begin{table}
\footnotesize
\caption{\label{table:features_DDoS2019_per_attack} CICDDoS2019 features importance per attack vector. A comparison between our model (packet-based features) and flow-based features importance as reported by \cite{sharafaldin2019developing}}
\center
    \begin{tabular}{m{1.9cm} m{1.2cm} m{2.2cm} m{1.1cm}}
    \hline
      \multicolumn{2}{c}{Our Model} & \multicolumn{2}{c}{CICDDoS2019 - Flow Records}\\
    \cline{1-4}
     Feature & Importance & Feature & Importance \\ \hline
        \multicolumn{4}{l}{UDP-Lag}\\ \hline
         is\_forward & 0.43834 & ACK Flag Count & 0.12544 \\ 
         is\_ACK\_flag & 0.42032 & Init Win bytes & 0.00209 \\ 
         length & 0.10657 & min seg size fwd & 0.00079 \\ 
         init\_win\_bytes & 0.03474 & Fwd IAT Mean & 0.00061 \\ 
         Timestamp & 2.9E-07 & Fwd IAT Max & 0.00047 \\ \hline
        \multicolumn{4}{l}{NetBIOS}\\ \hline
         length & 0.95687 & Fwd Packets/s & 0.00017 \\ 
         is\_RST\_flag & 0.03493 & min seg size fwd & 7.2E-05 \\ 
         init\_win\_bytes & 0.00302 & Protocol & 4.60E-05 \\ 
         Index & 0.00278 & \scriptsize{Fwd header length} & 3.5E-05 \\ 
         is\_forward & 0.00175 & N/A & N/A \\ \hline
         \multicolumn{4}{l}{SYN}\\ \hline
         length & 0.81002 & ACK Flag Count & 0.14583 \\
         is\_ACK\_flag & 0.17956 & \scriptsize{Init win bytes fwd} & 0.00243 \\
         direction\_iat\_after & 0.01041 & min seg size fwd & 0.00087 \\ 
         iat\_before & 1.0E-06 & fwd IAT Total & 0.00057 \\ 
         Timestamp & 3.3E-07 & Flow duration & 0.00041 \\ \hline
         \multicolumn{4}{l}{UDP}\\ \hline
         length & 0.93629 & Destination Port & 0.00070 \\
         iat\_after & 0.01645 & \scriptsize{fwd packet len. std} & 0.00061 \\ 
         init\_win\_bytes & 0.01513 & \scriptsize{Packet len. std} & 0.00024 \\ 
         Destination Port & 0.00871 & min seg size fwd & 9.8E-05 \\
         Timestamp & 0.00227 & Protocol & 4.5E-05 \\ \hline
        \multicolumn{4}{l}{MSSQL}\\ \hline
         length & 0.97563 & Fwd Packets/s & 0.00020 \\ 
         init\_win\_bytes & 0.01790 & Protocol & 4.6E-05 \\ 
         is\_forward & 0.00471 & N/A & N/A \\ 
         Destination port & 0.00124 & N/A & N/A \\ 
         Timestamp & 0.00030 & N/A & N/A \\ \hline         
         \multicolumn{4}{l}{LDAP}\\ \hline
         length & 0.97827 & \scriptsize{Max packet len} & 1.27832 \\ 
         is\_RST\_flag & 0.00981 & \scriptsize{Fwd packet len max} & 0.14322 \\ 
         init\_win\_bytes & 0.00456 & \scriptsize{Fwd packet len min} & 0.00873 \\
         Index & 0.00236 & avg packet size & 0.00653 \\ 
         Destination Port & 0.00151 & min packet len & 0.00391 \\ \hline
         \multicolumn{4}{l}{TFTP}\\ \hline
         is\_SYN\_flag& 0.95546 & Fwd IAT Mean & 0.00021 \\ 
         Destination Port & 0.04395 & min seg size fwd & 0.00019 \\ 
         is\_forward & 0.00055 & Fwd IAT Max & 0.00015 \\
         Timestamp & 3.2E-05 & Flow IAT Max & 0.00012 \\ 
         \hline  
         \multicolumn{4}{l}{SNMP}\\ \hline
         length & 0.98754 & \scriptsize{Max packet len} & 1.15205 \\ 
         is\_RST\_flag & 0.00754 & \scriptsize{Fwd packet len max} & 0.12907 \\ 
         Index & 0.00306 & \scriptsize{Fwd packet len min} & 0.00788 \\
         init\_win\_bytes & 0.00171 & avg packet size & 0.00591 \\
         \hline     
         \multicolumn{4}{l}{SSDP}\\ \hline
         length & 0.99999 & Destination Port & 0.00067 \\ 
         iat\_after,  & 2.7E-07 & \scriptsize{fwd packet len. std} &  0.00060\\ 
         N/A & N/A & \scriptsize{Packet len. std} & 0.00023 \\
         N/A & N/A & Protocol & 4.6E-05 \\ 
        \hline      
         \multicolumn{4}{l}{DNS}\\ \hline
         length & 0.94028 & \scriptsize{Max packet len} & 1.13986 \\ 
         iat\_after & 0.05929 & \scriptsize{Fwd packet len max} & 0.12771 \\ 
         is\_forward & 0.00023 & \scriptsize{Fwd packet len min} & 0.00779 \\
         init\_win\_bytes & 0.00019 & avg packet size & 0.00585 \\ 
        \hline   
        \multicolumn{4}{l}{NTP}\\ \hline
         is\_forward & 0.83114 & Subflow Fwd Bytes & 0.10648 \\ 
         length & 0.07857 & \scriptsize{Fwd Packets len} & 0.05802 \\ 
         Destination Port & 0.06857  & \scriptsize{Fwd Packet len std} & 0.00108 \\
         init\_win\_bytes & 0.02171 & min seg size fwd & 0.00070 \\ 
        \hline 
    \end{tabular}
\end{table}

\begin{table}
\footnotesize
\caption{\label{table:features_IDS2017_per_attack} CICIDS2017 features importance per attack vector. A comparison between our model (packet-based features) and flow-based features importance as reported by \cite{sharafaldin2018toward}}
\center
    \begin{tabular}{m{1.9cm} m{1.2cm} m{2.2cm} m{1.1cm}}
    \hline
      \multicolumn{2}{c}{Our Model} & \multicolumn{2}{c}{CICIDS2017 - Flow Records}\\
    \cline{1-4}
     Feature & Importance & Feature & Importance \\ \hline
        \multicolumn{4}{l}{DoS Goldeneye}\\ \hline
          iat\_before & 0.40140 & \scriptsize{Bwd Packet len std} & 0.1585 \\ 
          iat\_after & 0.18961 & Flow IAT Min & 0.0317 \\ 
          is\_RST\_flag & 0.14305 & Fwd IAT Min & 0.0257 \\ 
          init\_win\_bytes & 0.13187 & Flow IAT Mean & 0.0214 \\ \hline
        \multicolumn{4}{l}{DoS Hulk}\\ \hline
          length & 0.56913 & \scriptsize{Bwd Packet len std}  & 0.2028 \\ 
          is\_forward & 0.23272 & Flow Duration & 0.0437 \\ 
          is\_RST\_flag & 0.09473 & Flow IAT Std & 0.0227 \\ 
          iat\_before & 0.06688 & N/A & N/A  \\ \hline        
        \multicolumn{4}{l}{DoS Slowhttp}\\ \hline
          iat\_after & 0.36923 & Flow Duration & 0.0443 \\ 
          iat\_before & 0.34921 & Active Min & 0.0228 \\ 
          length & 0.08645 & Active Mean & 0.0219 \\ 
          Index &  0.07819 & Flow IAT Std & 0.0200 \\ \hline  
        \multicolumn{4}{l}{DoS slowloris}\\ \hline
          iat\_before & 0.48649 & Flow Duration & 0.0431 \\ 
          iat\_after & 0.15583 & Fwd IAT Min & 0.0378 \\ 
          length & 0.12665 & Bwd IAT Mean & 0.0300 \\ 
          is\_PSH\_flag & 0.12298 & Fwd IAT Mean & 0.0265 \\ \hline  
        \multicolumn{4}{l}{SSH-Patator}\\ \hline
          length & 0.96662 & Init win fwd bytes & 0.0079 \\ 
          is\_SYN\_flags & 0.03278 & Subflow FWd Bytes & 0.0052 \\ 
          is\_PSH\_flag & 0.00058 & Total fwd len & 0.0034 \\ 
          N/A & N/A & ACK Flag Count & 0.0007 \\ \hline  
        \multicolumn{4}{l}{FTP-Patator}\\ \hline
          Destination Port & 0.99745 & Init win fwd bytes & 0.0077 \\ 
          is\_PSH\_flag & 0.00237 & Fwd PSH Flags & 0.0062 \\ 
          init\_win\_bytes & 0.00016 & SYN Flag Count & 0.0061 \\ 
          N/A & N/A & Fwd Packets/s & 0.0014 \\ \hline  
        \multicolumn{4}{l}{Web Attack}\\ \hline
          is\_SYN\_flags & 0.68899 & Init win fwd bytes & 0.0200 \\ 
          iat\_before & 0.14606 & Subflow Fwd bytes & 0.0145 \\ 
          is\_PSH\_flag & 0.09172 & Init win bwd bytes  & 0.0129 \\ 
          Timestamp & 0.06940 & Total fwd len  & 0.0096 \\ \hline  
        \multicolumn{4}{l}{Bot}\\ \hline
          iat\_after &  0.99959 & Subflow Fwd bytes & 0.0239 \\ 
          iat\_before & 0.00041  & Total fwd len  & 0.0158 \\ 
          N/A & N/A & Mean fwd len & 0.0025 \\ 
          N/A & N/A & Bwd Packets/s & 0.0021 \\ \hline    
        \multicolumn{4}{l}{PortScan}\\ \hline
          is\_RST\_flag & 0.91366 & Init win fwd bytes & 0.0083 \\ 
          iat\_after & 0.07957 & Bwd Packets/s & 0.0032 \\ 
          length & 0.00651 & PSH Flag Count & 0.0009 \\ 
          iat\_before & 0.00011 & N/A & N/a  \\ \hline    
                 \multicolumn{4}{l}{DDoS
}\\ \hline
          is\_ACK\_flag & 0.61109 & Bwd Packet Len Std & 0.1728 \\ 
          iat\_after &  0.30980 & Avg Packet Size & 0.0162 \\ 
          Destination Port & 0.06172 & Flow Duration & 0.0137 \\ 
          length & 0.01046 & Flow IAT Std & 0.0086 \\ \hline   
    \end{tabular}
\end{table}

In addition to the classifiers that were trained and evaluated using multiple attack vectors on each one of the datasets, here we train additional classifiers, each of which is trained on a single attack vector within the CICDDoS2019 and CICIDS2017 datasets, in order to provide features importance insights for each attack vector.

For each attack vector in both datasets, alongside the obtained features importance of our tree-based classifiers, we provide a comparison to the features importance values that were reported by the datasets' authors \cite{sharafaldin2019developing}, \cite{sharafaldin2018toward}. 
The comparison results for CICDDoS2019 and CICIDS2017 appear in Tables \ref{table:features_DDoS2019_per_attack}, \ref{table:features_IDS2017_per_attack} respectively.

\end{document}